\documentclass{article}

\usepackage{arxiv}

\usepackage[utf8]{inputenc} 
\usepackage[T1]{fontenc}    
\usepackage{nicefrac}       
\usepackage{microtype}      
\usepackage{amsmath}
\usepackage{amssymb}
\usepackage{amsfonts}
\usepackage{amsthm}         

\usepackage{graphicx}
\usepackage{booktabs}       
\usepackage{multirow}
\usepackage[table]{xcolor}
\usepackage[caption=false,font=normalsize,labelfont=sf,textfont=sf]{subfig}

\usepackage[ruled,vlined,linesnumbered]{algorithm2e}
\SetKwInput{KwInput}{Input}
\SetKwInput{KwOutput}{Output}
\SetKw{KwInit}{Initialize}
\SetKw{KwRet}{return}
\SetKw{KwAnd}{and}
\SetKw{KwOr}{or}

\usepackage{url}          
\usepackage{natbib}
\let\cite\citep
\usepackage{doi}
\usepackage{hyperref}       

\usepackage[shortcuts,acronym]{glossaries}
\glsdisablehyper
\newif\ifuniqueAffiliation
\uniqueAffiliationfalse

\usepackage{authblk}

\newbox{\orcid}\sbox{\orcid}{\includegraphics[scale=0.06]{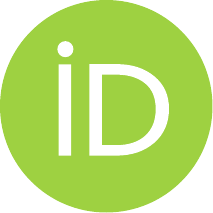}} 

\newacronym{ML}{ML}{machine learning}
\newacronym{DL}{DL}{deep learning}
\newacronym{MnL}{MnL}{manifold learning}
\newacronym{AE}{AE}{autoencoder}

\title{Explainable Autoencoder-Based Anomaly Detection in IEC 61850 GOOSE Networks \thanks{This work has been submitted to the IEEE for possible publication}}

\date{}

\author[1]{%
    \href{https://orcid.org/0000-0002-4521-0031}{\usebox{\orcid}\hspace{1mm}Dafne Lozano-Paredes\thanks{\texttt{dafne.lozanop@urjc.es}}}%
}
\author[1]{%
    \href{https://orcid.org/0000-0001-8845-2834}{\usebox{\orcid}\hspace{1mm}{Luis Bote-Curiel}%
}}

\author[3]{%
    \href{https://orcid.org/0000-0001-9685-4032}{\usebox{\orcid}\hspace{1mm}{Juan Ramón Feijoo-Martínez}%
}}

\author[1]{%
    \href{https://orcid.org/0000-0003-4673-8193}{\usebox{\orcid}\hspace{1mm}{Ismael Gómez-Talal}%
}}

\author[1,2]{%
  \href{https://orcid.org/0000-0003-0426-8912}{%
    \mbox{\usebox{\orcid}\hspace{1mm}José Luis Rojo-Álvarez}%
  }%
}

\affil[1]{Departamento de Teoría de la Señal y Comunicaciones y Sistemas Telemáticos y Computación, \protect\\ Universidad Rey Juan Carlos, Madrid, Spain.}
\affil[2]{Centro de Investigación for Data, Complex Networks and Cybersecurity Sciences,  \protect\\ Universidad Rey Juan Carlos, Madrid, Spain.}
\affil[3]{Red Eléctrica (REDEIA Group), \protect\\ Alcobendas, Madrid, Spain.}
\renewcommand{\shorttitle}{Explainable Autoencoder Anomaly Detection in GOOSE Networks}

\hypersetup{
pdftitle={Explainable Autoencoder-Based Anomaly Detection in IEC 61850 GOOSE Networks},
pdfsubject={cs.CR, cs.LG},
pdfauthor={Lozano-Paredes et al.},
pdfkeywords={Anomaly detection, Autoencoders, GOOSE protocol, IEC 61850, industrial control systems, intrusion detection systems, interpretability, manifold learning.}
}

\begin{document}
\maketitle

\begin{abstract}
The IEC 61850 Generic Object-Oriented Substation Event (GOOSE) protocol plays a critical role in real-time protection and automation of digital substations, yet its lack of native security mechanisms can expose power systems to sophisticated cyberattacks. Traditional rule-based and supervised intrusion detection techniques struggle to detect protocol-compliant and zero-day attacks under significant class imbalance and limited availability of labeled data. This paper proposes an explainable, unsupervised multi-view anomaly detection framework for IEC 61850 GOOSE networks that explicitly separates semantic integrity and temporal availability. The approach employs asymmetric autoencoders trained only on real operational GOOSE traffic to learn distinct latent representations of sequence-based protocol semantics and timing-related transmission dynamics in normal traffic. Anomaly detection is implemented using reconstruction errors mixed with statistically grounded thresholds, enabling robust detection without specified attack types. Feature-level reconstruction analysis provides intrinsic explainability by directly linking detection outcomes to IEC 61850 protocol characteristics. The proposed framework is evaluated using real substation traffic for training and a public dataset containing normal traffic and message suppression, data manipulation, and denial-of-service attacks for testing. Experimental results show attack detection rates above 99\% with false positives remaining below 5\% of total traffic, demonstrating strong generalization across environments and effective operation under extreme class imbalance and interpretable anomaly attribution.
\end{abstract}



\section{Introduction}
\label{sec:introduction}
Extensive network communication capabilities have been integrated into critical infrastructure as a result of the modernization of electrical power systems into smart grids. \cite{Hussain2021, Achaal2024}. The integration of Operational Technology (OT) with advanced measurement and distributed control has significantly expanded the attack surface of digital substations \cite{Bindra2017}. Because successful intrusions can endanger system availability and integrity and even result in disastrous outcomes like equipment destruction or widespread outages, cybersecurity has become a top priority \cite{Hussain2021, Bindra2017, Achaal2024}. The Generic Object-Oriented Substation Event (GOOSE) protocol is important to the digital substation architecture specified by the IEC 61850 standard \cite{Mackiewicz2006}. GOOSE messages are essential for automation and real-time protection because they allow Intelligent Electronic Devices (IEDs) to communicate vital information \cite{Mackiewicz2006, Asim2020}. The standard requires communication times to be within milliseconds in order to guarantee the efficacy of electrical protection \cite{Mackiewicz2006}.  A simplified architecture that avoids the TCP/IP stack and maps straight to the Ethernet layer is required due to this strict latency requirement. However, this design decision results in a crucial trade-off because speed is given precedence over security, leading to a protocol without integrated encryption or authentication. \cite{Hussain2020, ABDELKADER2024, Youssef2016}.

This lack of inherent security exposes GOOSE traffic to sophisticated cyberattacks \cite{Hoyos2012}. Adversaries can exploit the multicast nature of GOOSE to launch Message Suppression (MS) \cite{Wright018}, Data Manipulation (DM) \cite{HABIB20238}, or complex masquerade attacks \cite{Wang2022}, where the attacker mimics legitimate protocol behavior by manipulating the Status Number (\texttt{stNum}) and Sequence Number (\texttt{sqNum}) \cite{Hussain2020, Elbez2024}. Traditional signature-based intrusion detection systems have difficulty detecting zero-day attacks and subtle manipulations that occur in dynamic industrial control systems \cite{Elbez2024, Petr2020}. To address these limitations, recent research has increasingly focused on anomaly detection systems based on Machine Learning (ML) and Deep Learning (DL) \cite{Nhung-Nguyen2024, Li2023}. Unsupervised models, particularly Autoencoders (AEs), have shown promise by learning the structure of benign traffic and identifying anomalies through reconstruction error \cite{Sakurada2014}. Supervised approaches often achieve high accuracy but depend on large volumes of labeled data containing known attack patterns, which are typically scarce or unavailable in operational OT environments \cite{DEOLIVEIRA2025}. As a result, supervised models often struggle to adapt to changes in network behavior and fail to generalize to zero-day attacks or previously unseen anomalies \cite{JAIN2025, Dai2024}.

In this work, we address these limitations by explicitly dividing semantic integrity and temporal availability through a multi-view anomaly detection framework based on different architectures of AEs trained exclusively on real operational GOOSE traffic. By learning separate representations for protocol-level sequencing behavior and timing-related dynamics, the proposed approach enables robust detection under severe class imbalance while remaining independent of labeled attack data or predefined attack patterns. Moreover, feature-level reconstruction analysis provides interpretability insights directly aligned with IEC 61850 protocol characteristics, supporting post-event analysis in practical substation environments.

The remainder of this paper is organized as follows. Section~\ref{sec:related_work} reviews the background and related work on anomaly detection for IEC~61850 GOOSE communication. Section~\ref{sec:Datasets} describes the datasets used in this study, and Section~\ref{sec:FeaturesEng} presents the feature engineering process. Section~\ref{sec:methodology} details the proposed methodology. Section~\ref{sec:results} presents the experimental setup and the obtained results. Finally, Section~\ref{sec:discussion} discusses the findings, and Section~\ref{sec:conclusion} concludes the paper.

\section{Background and Related Work}
\label{sec:related_work}

As discussed earlier, IEC~61850 has become the standard communication framework for digital substations, enabling interoperable protection, control, and monitoring among IEDs. Within this framework, GOOSE protocol is used to transmit time-critical status changes and trip signals over Ethernet multicast \cite{Mackiewicz2006, Asim2020}. Due to its strict latency requirements, GOOSE was designed without native authentication or encryption mechanisms, a limitation that has long been recognized as a security concern \cite{Hoyos2012}. Studies showed that a capable network adversary can create malicious packets that cause unwanted actions by taking advantage of GOOSE traffic multicast nature and predictable behavior. These findings indicate that protocol-compliant attacks can have direct physical repercussions in real substations \cite{Reda2021, Rajkumar2020}.

Research on anomaly detection for IEC~61850-based substations has evolved from simple rule-based techniques toward more advanced ML and DL approaches. This evolution reflects the need to overcome the limitations of static rules while still accounting for the fast and deterministic characteristics of GOOSE communication.

\subsection{Rule-Based and Statistical Approaches}

Several studies have exploited the deterministic behavior of GOOSE application-layer counters as a basis for rule-based intrusion detection. In these approaches, unexpected decreases or discontinuities in the retransmission sequences are treated as anomalies, leading to the immediate rejection of the affected packets and subsequent inspection of the message content to identify potential masquerading behavior \cite{Premaratne2010, Hussain2023}. Building on these counter-based rules, later work has proposed more comprehensive adversary models, extending the same sequencing and timing assumptions to both GOOSE and Sampled Values traffic. These models combine counter consistency checks with retransmission timing constraints to define rule-based detection mechanisms for replay and masquerade attacks \cite{Nweke2020}. Beyond specification-based rules, statistical modeling techniques have also been explored to detect deviations in GOOSE traffic behavior. Methods based on Auto-Regressive Fractionally Integrated Moving Average (ARFIMA) models characterize the long-memory dynamics of network traffic under normal conditions. Significant deviations between observed traffic and the learned statistical model are interpreted as indicators of flooding attacks \cite{Elbez2020}. While these deterministic approaches offer low false-positive rates for known attacks, they lack the flexibility to generalize to novel, hidden attacks that mimic legitimate traffic patterns or operate within accepted specifications but carry malicious intent \cite{Ustun2021}.

\subsection{Supervised Machine Learning}

To overcome the inflexibility of rule-based systems, ML algorithms have been applied to detect attacks that adhere to protocol specifications but exhibit statistical anomalies. Unlike deterministic rules, ML classifiers such as Support Vector Machines (SVM), Random Forest (RF), and Decision Trees operate on high-dimensional feature sets extracted from GOOSE headers and payloads, including inter-arrival times and sequence differentials \cite{Ahanger2021, Ustun2021a, Nhung-Nguyen2024}. Research demonstrates that ensemble methods achieve superior classification accuracy by aggregating decision boundaries from multiple learners, effectively reducing the false-positive rates common in single-threshold systems \cite{Ustun2021}. Building on this, optimizations such as Proximal Support Vector Machines have been used to maximize detection performance while reducing computational cost \cite{Hasmat2022}. The feasibility of these algorithms has been further validated in comparable real-time industrial use cases, confirming their applicability under the strict latency constraints required by critical infrastructure \cite{Saghezchi2022}. Moreover, to address the challenge of imbalanced datasets, recent studies have adopted traffic augmentation and data balancing techniques, enabling the successful training of time-series compatible ML algorithms with high sensitivity to scarce attack samples\cite{Bhattacharya2024}.

\subsection{Deep Learning and Autoencoder-based Anomaly Detection}

While traditional ML models are more flexible than strict rules, they often struggle when dealing with complex raw data. Because of this, recent research has shifted toward DL architectures, which can find attacks more effectively by learning the features automatically \cite{DEOLIVEIRA2025, Ferrag2020}. One popular method involves Autoencoders (AEs) which are designed to learn the structure and evolution of normal traffic. In this framework, the model is trained exclusively on benign traffic to learn a compressed latent representation of normal network dynamics. During inference, the system attempts to reconstruct the input from this learned feature space; since the model has not captured the statistical patterns of malicious traffic, it fails to accurately reconstruct anomalous packets, resulting in a significant reconstruction error that is used as the detection threshold \cite{Trinal2025, Jay2022}. Other approaches use the discrete wavelet transform to extract deep details from the data and combine it with locally linear embedding to simplify those details. These optimized features are then passed to a Long Short-Term Memory (LSTM)-based AE that reconstructs the time-based patterns, allowing for fast and accurate detection in real substations \cite{Yang2022}.

Beyond these unsupervised methods, researchers have also developed powerful supervised models to target specific threats. For instance, Deep Convolutional Neural Networks (DCNN) using GPU acceleration have shown they can identify attacks with over 99\% accuracy on large public datasets \cite{Hnamte2023}. Furthermore, hybrid architectures that combine DCNN with LSTM units have been built. Unlike standard detection models, these systems use CNNs to extract spatial features from measurement data and LSTMs to predict missing temporal sequences, effectively compensating for data lost during attacks. This reconstruction capability maintains the accuracy of power system state estimation even when the communication network is under active interference \cite{Xu2022}.

\section{Datasets}
\label{sec:Datasets}

This work relies on a combination of a real-world dataset captured in an operational IEC~61850 environment and a publicly available dataset containing cyberattack scenarios. Using both sources allows us to model legitimate GOOSE traffic while evaluating the generalization of the anomaly detector under controlled attacks. This hybrid strategy avoids overfitting to laboratory conditions and enables assessment of the model ability to generalize across different environments.

\subsection{Real-World GOOSE Dataset from an IEC~61850 Substation}

To characterize the normal behavior of GOOSE traffic, packet captures were collected from an operational IEC~61850 LAN inside a high-voltage substation from the Red Eléctrica (REDEIA Group) telecommunication network. Traffic was recorded passively using a network tap or a switch SPAN port employed for routine monitoring, where multiple IEDs exchanged protection and status information using GOOSE, along with other IEC~61850-related services such as Parallel Redundancy Protocol (PRP) and High-availability Seamless Redundancy (HSR) supervision, as well as Link Layer Discovery Protocol (LLDP).

All captures correspond to steady-state operation: no switching events, protection triggers, or injected disturbances occurred during acquisition. The traces therefore reflect the natural background of an operational substation, including periodic GOOSE retransmissions, synchronization messages, and low-rate management traffic. Several GOOSE publishers appear in the dataset, each with its own retransmission intervals or event-driven behavior. Packet capture (PCAP) files contain several thousand GOOSE frames, which are subsequently aggregated into sliding windows for feature extraction.

This real-world dataset is used here exclusively for training the AE, reflecting the common situation in substations where only normal traffic is available.

\subsection{Public IEC~61850 Dataset with Attack Scenarios}

The detection performance of the model is evaluated using the IEC61850SecurityDataset \cite{Biswas2019}, a publicly available collection of GOOSE traces recorded in a laboratory substation. The setup comprises 18 IEDs arranged in a realistic four-bus substation topology. The dataset includes normal operation, physical disturbance events (e.g., busbar faults, breaker failures, under-frequency load shedding), and three representative classes of cyberattacks:

1) \emph{MS} are attacks that interfere with the normal propagation of state changes by disrupting the expected evolution of GOOSE sequence counters, resulting in missing or inconsistent state updates at subscribing IEDs~\cite{Wright018}; 
2) \emph{DM} are attacks that alter the semantic content of GOOSE messages by forging control or status information, such as breaker positions or measurement values, while preserving protocol-compliant structure~\cite{HABIB20238}; and 
3) \emph{Denial-of-Service (DoS)} are availability attacks that inject high volumes of false GOOSE frames, overwhelming subscribers and disrupting the timely processing of legitimate messages~\cite{Wang2022}.

Each scenario is provided as an approximately 10-minute PCAP trace, generated by injecting malicious frames into an otherwise normal GOOSE traffic sequence. This dataset, which includes labeled attack instances, is used here exclusively for testing and performance evaluation.

\section{Feature Engineering: Temporal and Sequence Dynamics of GOOSE}
\label{sec:FeaturesEng}

GOOSE traffic exhibits very consistent timing patterns and deterministic application-layer sequencing rules. To capture these complementary aspects, features are computed over sliding windows ($T_w$) of 0.1, 0.5, 1, and 3~s. Packets are first grouped into flows, defined by the combination of GOOSE identification and the source Medium Access Control (MAC) address, then ordered by timestamp and summarized within each window.

All features are derived from IEC~61850 semantics and the rules governing GOOSE retransmissions. Rather than treating the traffic as a homogeneous signal, the feature set is intentionally decomposed into two complementary domains that reflect distinct behavioral modalities: temporal dynamics and sequence semantics.

Temporal features describe the continuous timing and volumetric behavior of GOOSE traffic and capture deviations related to availability and transmission patterns. As summarized in Table~\ref{tab:TemporalF}, these features characterize inter-arrival times, packet rates, jitter, frame sizes, and protocol timing fields. Under normal operation, GOOSE retransmissions follow well-defined timing profiles, with limited variability across windows. Attacks that suppress, inject, or flood messages disrupt these patterns.

\begin{table*}[h]
\centering
\caption{Temporal and volumetric features extracted from GOOSE network traces.}
\label{tab:TemporalF}
\renewcommand{\arraystretch}{1.2} 
\begin{tabular}{p{3.3cm} p{7.2cm} p{6.3cm}}
\hline
\textbf{Feature} & \textbf{Description} & \textbf{Indicative Anomalous Behavior} \\
\hline
\textbf{dt} & Inter-arrival time between consecutive GOOSE frames within the same flow. & Not used directly for detection; serves as the basis for windowed temporal statistics. \\

\textbf{dt\_mean}, \textbf{dt\_std} & Mean and standard deviation of inter-arrival times within the sliding window $T_w$. & Deviations from expected timing profiles reveal missing messages, micro-bursts, altered retransmission behavior, or injection activity. \\

\textbf{rate\_mean} & Average packet transmission rate computed as the inverse of inter-arrival time. & Abnormally high rates indicate injection or flooding; sustained drops suggest message suppression or delivery disruption. \\

\textbf{pkt\_count} & Total number of GOOSE frames observed within the time window $T_w$. & Elevated counts signal volumetric DoS behavior; unusually low counts indicate suppressed communication. \\

\textbf{jitter} & Short-term variation in packet arrival times relative to a local moving median. & Not used directly for detection; serves as the basis for windowed jitter statistics. \\

\textbf{jitter\_mean}, \textbf{jitter\_std} & Mean and standard deviation of jitter values within the sliding window $T_w$. & Elevated jitter variability indicates congestion, queuing effects, or interference caused by injected or flooding traffic. \\

\textbf{len\_mean} & Average Ethernet frame length within the window. & Semantic payload manipulation can alter the encoding size, resulting in systematic changes in frame length. \\

\textbf{ttl\_mean} & Mean value of the timeAllowedToLive (TTL) parameter, which specifies the maximum validity duration of a GOOSE message before it is considered stale. & Inconsistent TTL values may indicate forged or replayed messages that are not synchronized with legitimate publisher behavior. \\

\hline
\end{tabular}
\end{table*}

Sequence-based features, summarized in Table~\ref{tab:SequenceF}, capture the evolution of application-layer counters and configuration identifiers embedded in GOOSE messages. The fields \texttt{stNum} and \texttt{sqNum} follow strict deterministic rules under legitimate operation: \texttt{stNum} changes only upon state transitions and \texttt{sqNum} increases monotonically with each retransmission. Deviations from these rules provide strong indicators of semantic integrity disruptions, including forged events, message suppression, and protocol-compliant manipulation.

\begin{table*}[!h]
\centering
\caption{Sequence-based features extracted from GOOSE network traces.}
\label{tab:SequenceF}
\renewcommand{\arraystretch}{1.2}
\begin{tabular}{p{3.3cm} p{7.2cm} p{6.3cm}}
\hline
\textbf{Feature} & \textbf{Description} & \textbf{Indicative Anomalous Behavior} \\
\hline
\textbf{st\_changes} & Number of state number (\textit{stNum}) increments observed within the window. & Unexpected or frequent state changes may indicate forged events or manipulation of control semantics. \\

\textbf{sq\_resets} & Count of decreases in the sequence number (\textit{sqNum}) within the window. & Non-monotonic sequence evolution indicates out-of-order delivery or injection of counterfeit messages. \\

\textbf{sq\_bigjump} & Number of instances where the increment of \textit{sqNum} exceeds one between consecutive frames. & Large jumps suggest missing, overwritten, or intentionally suppressed GOOSE frames. \\

\textbf{sq\_progress} & Range of \textit{sqNum} values observed within the window. & Legitimate traffic evolves smoothly; reduced or irregular progression reflects disrupted or manipulated sequencing. \\

\textbf{st\_jump\_size\_max} & Maximum observed increment between consecutive \textit{stNum} values. & Large state jumps indicate unlikely transitions inconsistent with legitimate process behavior. \\

\textbf{bad\_dst\_rate} & Fraction of GOOSE frames sent to non-multicast destination addresses. & Incorrect destination addressing is indicative of malformed or counterfeit traffic that does not conform to IEC~61850 specifications. \\
\hline
\end{tabular}
\end{table*}

After feature extraction, the windowed dataset is preprocessed to obtain a consistent numeric matrix for model training. Degenerate variables and windows without usable information are removed, while missing values are handled through flow-consistent interpolation for continuous features and zero-filling for event-based variables. Identifiers, timestamps, and labels are discarded, retaining only the feature set $\mathcal{F}$. The resulting data form a matrix $\mathbf{X} \in \mathbb{R}^{N \times |\mathcal{F}|}$, where $N$ denotes the number of valid sliding windows.

\section{Proposed Methodology}
\label{sec:methodology}
\begin{figure}[t]
    \centering
    \includegraphics[width=0.9\textwidth]{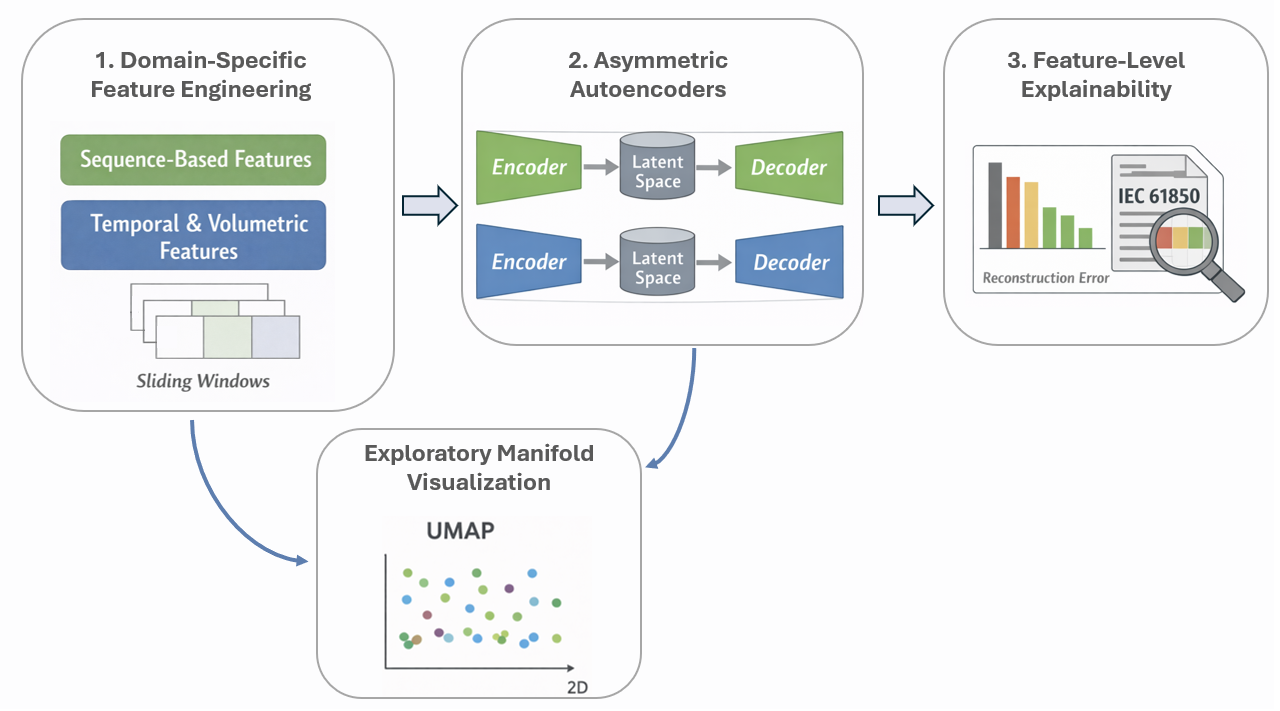}
    \caption{Overview of the proposed multi-view anomaly detection framework for IEC~61850 GOOSE communication. }
    \label{fig:pipeline}
\end{figure}

This section presents the proposed multi-view anomaly detection framework for IEC~61850 GOOSE traffic. As illustrated in Fig.~\ref{fig:pipeline}, the framework is structured into three main stages: domain-specific feature engineering, asymmetric latent representation learning with dynamic thresholding, and feature-level explainability to support post-event analysis. 

In addition, exploratory visualization of the raw feature space and learned latent representations is employed to analyze the underlying structure of the data and to support qualitative interpretation of detection behavior.

\subsection{Overview of the Multi-View Detection Framework}

Starting from the preprocessed feature matrix $\mathbf{X} \in \mathbb{R}^{N \times |\mathcal{F}|}$, where each row corresponds to a sliding window and $N$ denotes the number of valid windows, the proposed methodology models GOOSE traffic through two complementary views, namely, \emph{semantic integrity} and \emph{temporal availability}.

The feature set $\mathcal{F}$ is decomposed into two disjoint subsets,
$\mathcal{F} = \mathcal{F}_{\text{seq}} \cup \mathcal{F}_{\text{temp}}, \quad \text{and} \quad 
\mathcal{F}_{\text{seq}} \cap \mathcal{F}_{\text{temp}} = \emptyset,$
where $\mathcal{F}_{\text{seq}}$ contains sequence-based features derived from IEC~61850 application-layer semantics, explained in Table \ref{tab:SequenceF} and $\mathcal{F}_{\text{temp}}$ contains temporal and volumetric features describing transmission dynamics, explained in Table \ref{tab:TemporalF}.

Accordingly, each windowed observation $\mathbf{x}_i \in \mathbb{R}^{|\mathcal{F}|}$ is split into two subvectors, $\mathbf{x}_i^{\text{seq}} \in \mathbb{R}^{|\mathcal{F}_{\text{seq}}|}$ and $\mathbf{x}_i^{\text{temp}} \in \mathbb{R}^{|\mathcal{F}_{\text{temp}}|}$.

\subsection{Exploratory Manifold Analysis of the Raw Feature Space}

Prior to representation learning, the intrinsic structure of the raw feature space is analyzed to assess separability and variability across datasets and attack scenarios. For this purpose, Uniform Manifold Approximation and Projection (UMAP) \cite{McInnes2018} is applied independently to the sequence-based and temporal feature matrices, denoted as $\mathbf{X}^{\text{seq}}$ and $\mathbf{X}^{\text{temp}}$, respectively.

Formally, UMAP defines a nonlinear mapping from the high-dimensional feature space to a two-dimensional embedding:
\begin{equation}
\mathbf{X}^{v} \in \mathbb{R}^{N \times |\mathcal{F}_{v}|},
\quad
\mathbf{Y}^{v} \in \mathbb{R}^{N \times 2},
\quad
v \in \{\text{seq}, \text{temp}\}.
\end{equation}

Here, $\mathbf{X}^{v}$ represents the original high-dimensional feature matrix for view $v$, while $\mathbf{Y}^{v}$ denotes its corresponding two-dimensional embedding. This analysis is used exclusively as an exploratory and diagnostic tool and does not influence the detection process. By visualizing the low-dimensional embeddings of the original features, the degree of overlap between normal and anomalous behavior can be assessed, motivating the use of reconstruction-based anomaly detection.

\subsection{Asymmetric Autoencoder Models for Multi-View Representation Learning}

To model each behavioral domain independently, two AEs are trained using only normal traffic from the real-world dataset: a semantic AE for sequence-based features and a temporal AE for timing-related features.

For each view $v \in \{\text{seq}, \text{temp}\}$, the AE consists of an encoder
\begin{equation}
f_{\boldsymbol{\theta}_v} : \mathbb{R}^{|\mathcal{F}_v|} \rightarrow \mathbb{R}^{d_v}
\end{equation}
and a decoder
\begin{equation}
g_{\boldsymbol{\phi}_v} : \mathbb{R}^{d_v} \rightarrow \mathbb{R}^{|\mathcal{F}_v|},
\end{equation}
where $d_v$ denotes the dimensionality of the bottleneck layer.
Given an input sample $\mathbf{x}_i^{v}$, the reconstructed output is
\begin{equation}
\hat{\mathbf{x}}_i^{v} = g_{\boldsymbol{\phi}_v}
\bigl( f_{\boldsymbol{\theta}_v}(\mathbf{x}_i^{v}) \bigr).
\end{equation}

The AEs are intentionally asymmetric. The semantic model employs a deeper architecture to capture nonlinear protocol dependencies, while the temporal model uses a narrower bottleneck to prevent reconstruction of volumetric flooding behavior.

\subsection{Reconstruction-Based Anomaly Scoring and EVT Thresholding}

For each view, the reconstruction error is computed as the mean squared error,
\begin{equation}
e_i^{v} = \frac{1}{|\mathcal{F}_v|}
\left\| \mathbf{x}_i^{v} - \hat{\mathbf{x}}_i^{v} \right\|_2^2.
\end{equation}

Rather than relying on fixed or percentile-based thresholds, anomaly decisions are derived using Extreme Value Theory (EVT). The tail of the reconstruction error distribution obtained from normal training data is modeled using a generalized Pareto distribution, yielding a statistically grounded threshold $z_v^\ast$ for each view.

A window $i$ is flagged as anomalous if
\begin{equation}
A_i =
\mathbb{I}\!\left(e_i^{\text{seq}} > z_{\text{seq}}^\ast \right)
\;\lor\;
\mathbb{I}\!\left(e_i^{\text{temp}} > z_{\text{temp}}^\ast \right),
\end{equation}
where $\mathbb{I}(\cdot)$ denotes the indicator function.

\subsection{Explainability via Feature-Level Reconstruction Error}

To support interpretability and post-event analysis, the anomaly score is decomposed into per-feature contributions. For a given view $v$, the contribution of feature $j$ for window $i$ is computed as
\begin{equation}
c_{i,j}^{v} = \left(x_{i,j}^{v} - \hat{x}_{i,j}^{v}\right)^2.
\end{equation}

These feature-level contributions provide a direct mapping between detection outcomes and IEC~61850 protocol semantics, enabling intuitive interpretation of anomalies without relying on surrogate explainability models.

\section{Results}
\label{sec:results}

\subsection{Manifold Structure and Topological Characterization of the Feature Space}

\begin{figure*}[t]
    \centering
    \begin{tabular}{cccc}
        \multicolumn{4}{c}{\textbf{Semantic latent space (SEQ)}} \\
        \includegraphics[width=0.23\textwidth]{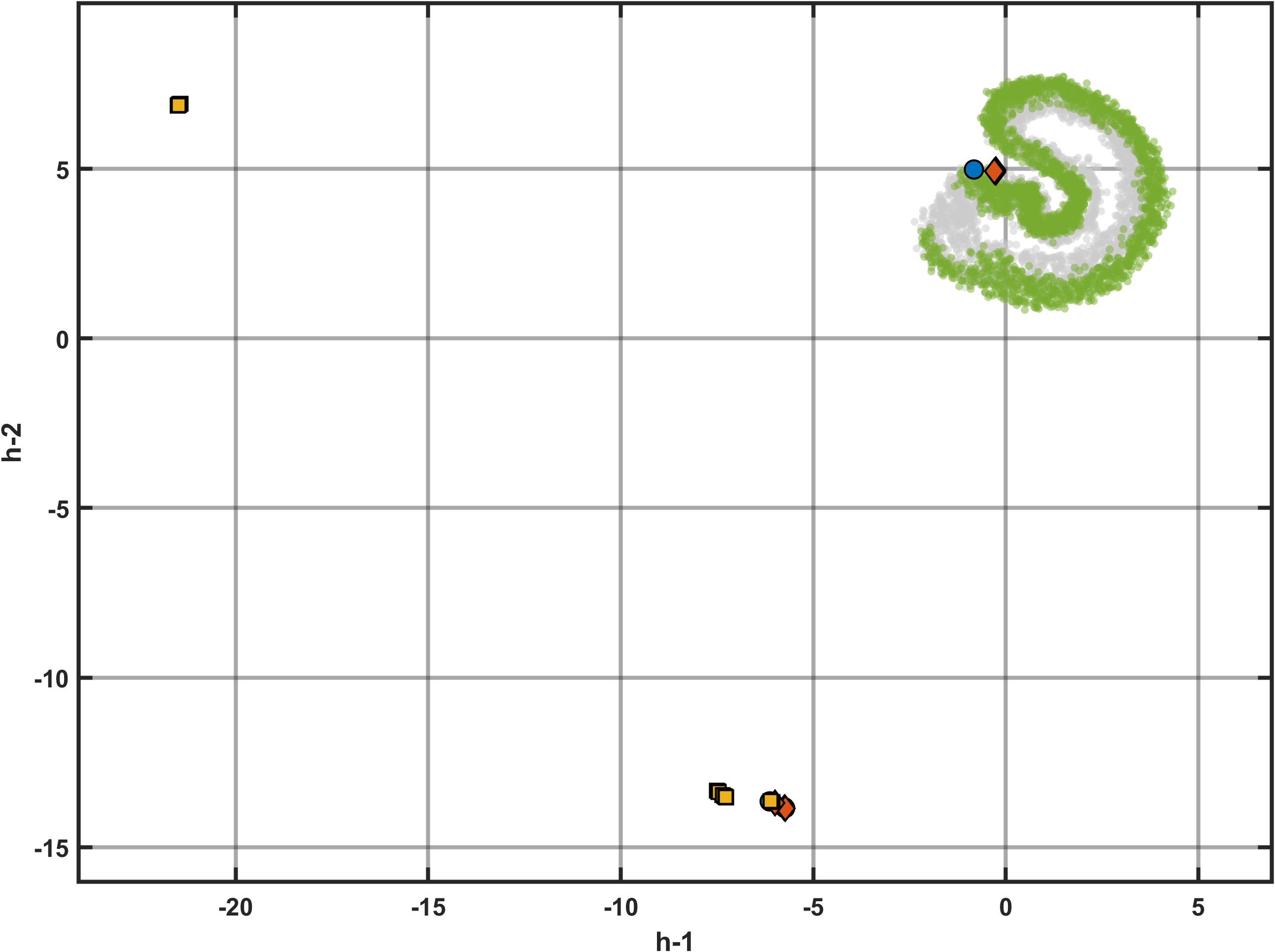} &
        \includegraphics[width=0.23\textwidth]{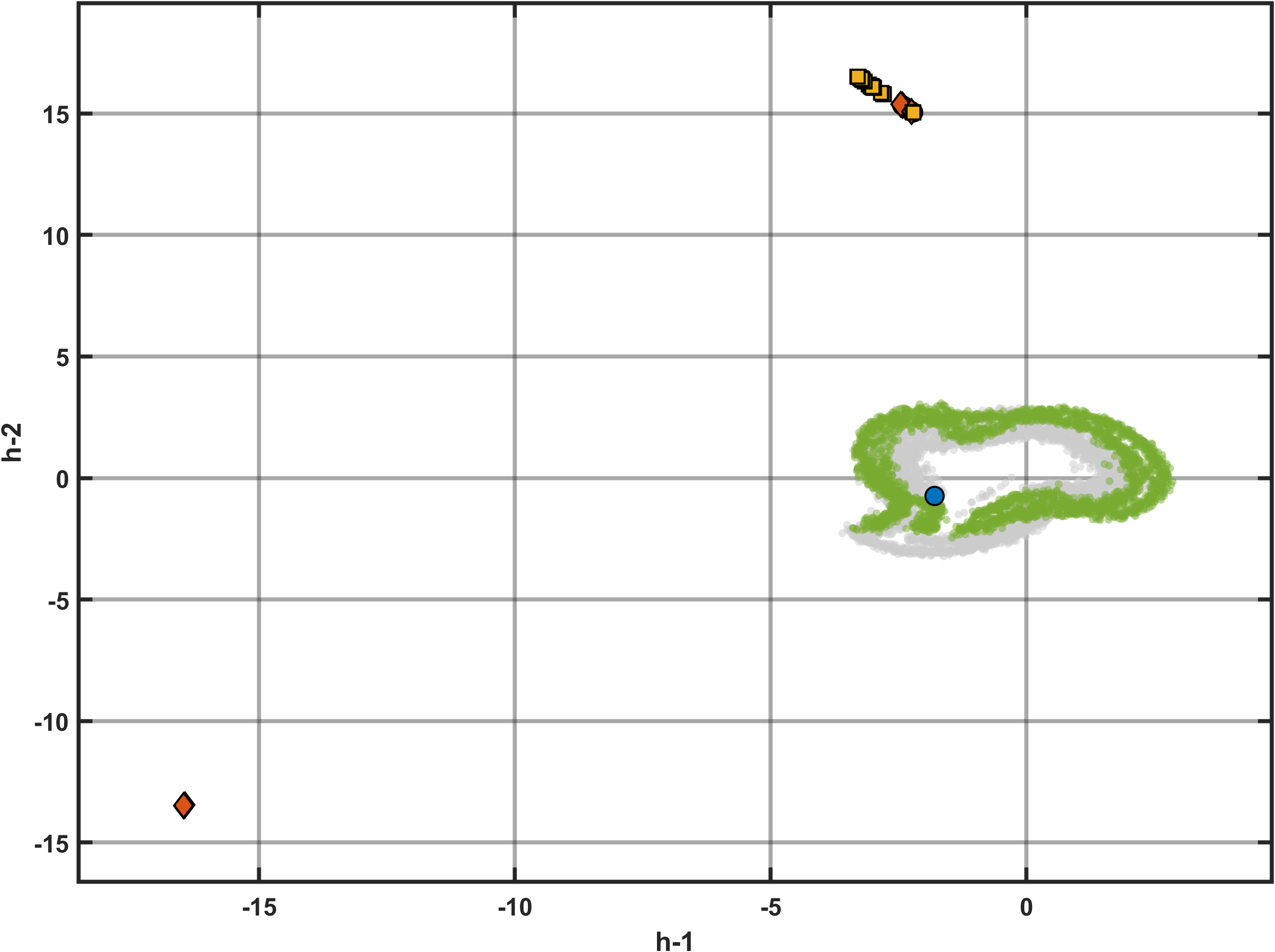} &
        \includegraphics[width=0.23\textwidth]{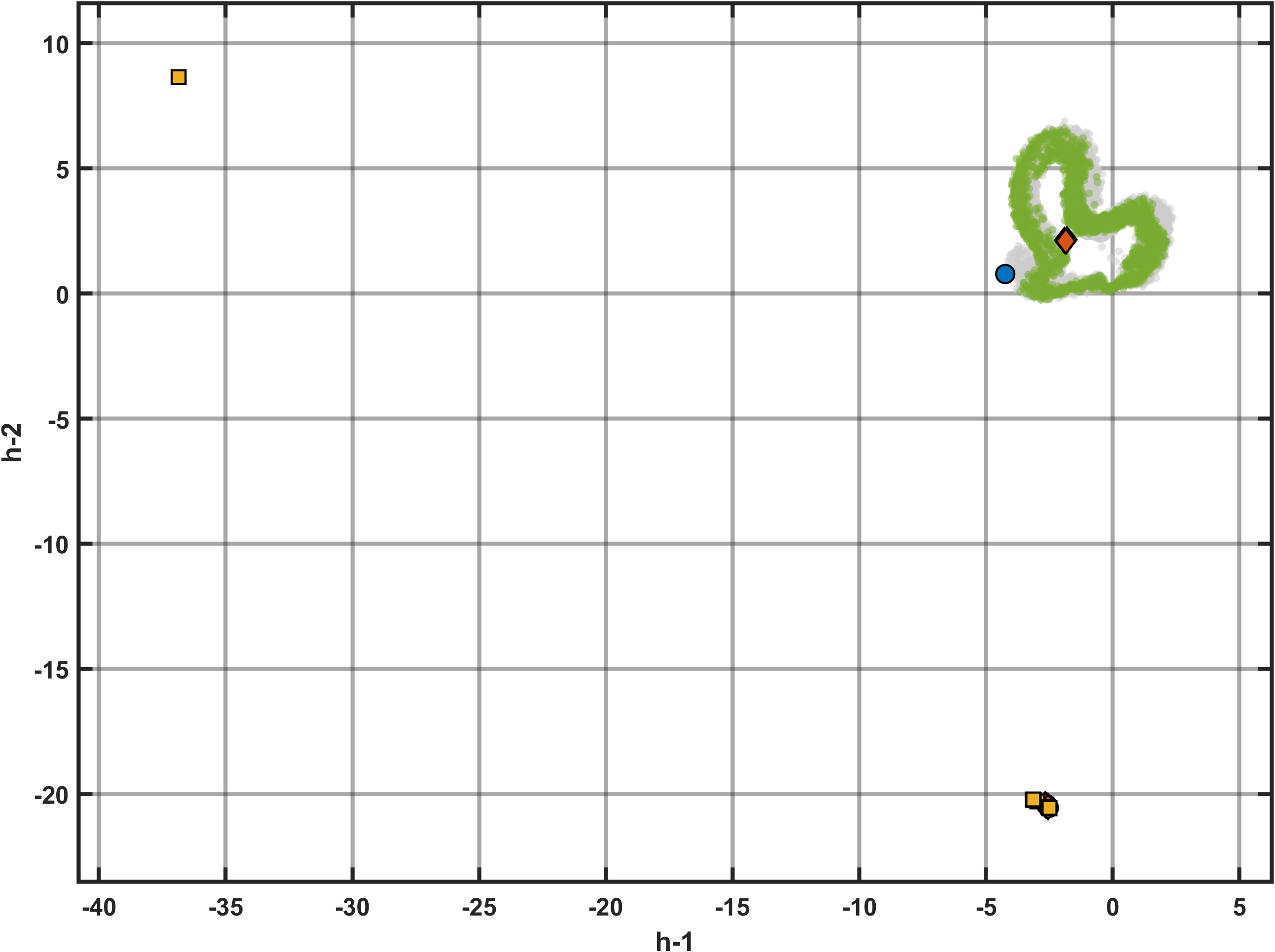} &
        \includegraphics[width=0.23\textwidth]{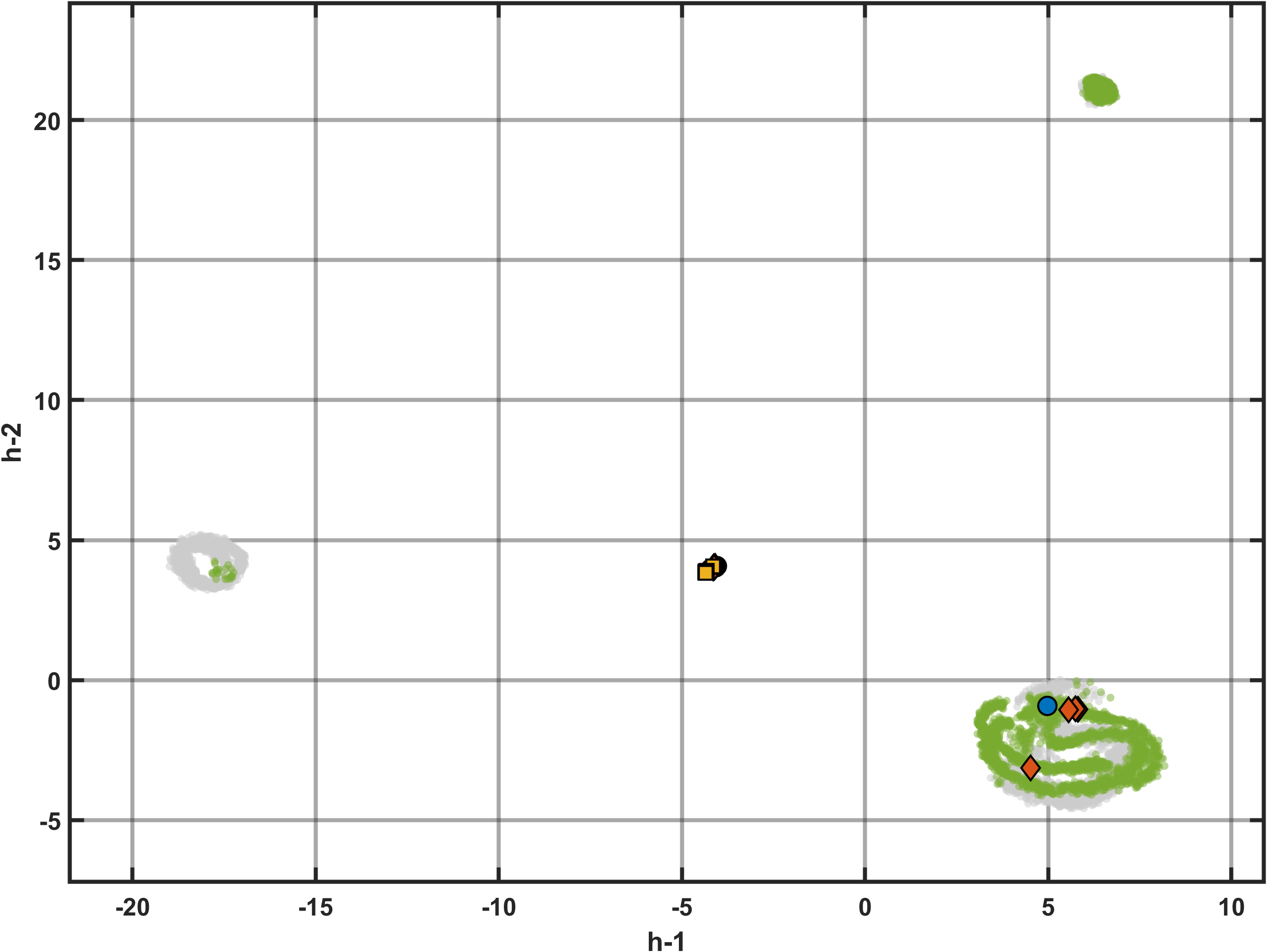} \\
        (a) $T_w = 0.1$ s & (b) $T_w = 0.5$ s & (c) $T_w = 1.0$ s & (d) $T_w = 3.0$ s \\
        [1.5ex]
        \multicolumn{4}{c}{\textbf{Temporal latent space (TEMP)}} \\
        \includegraphics[width=0.23\textwidth]{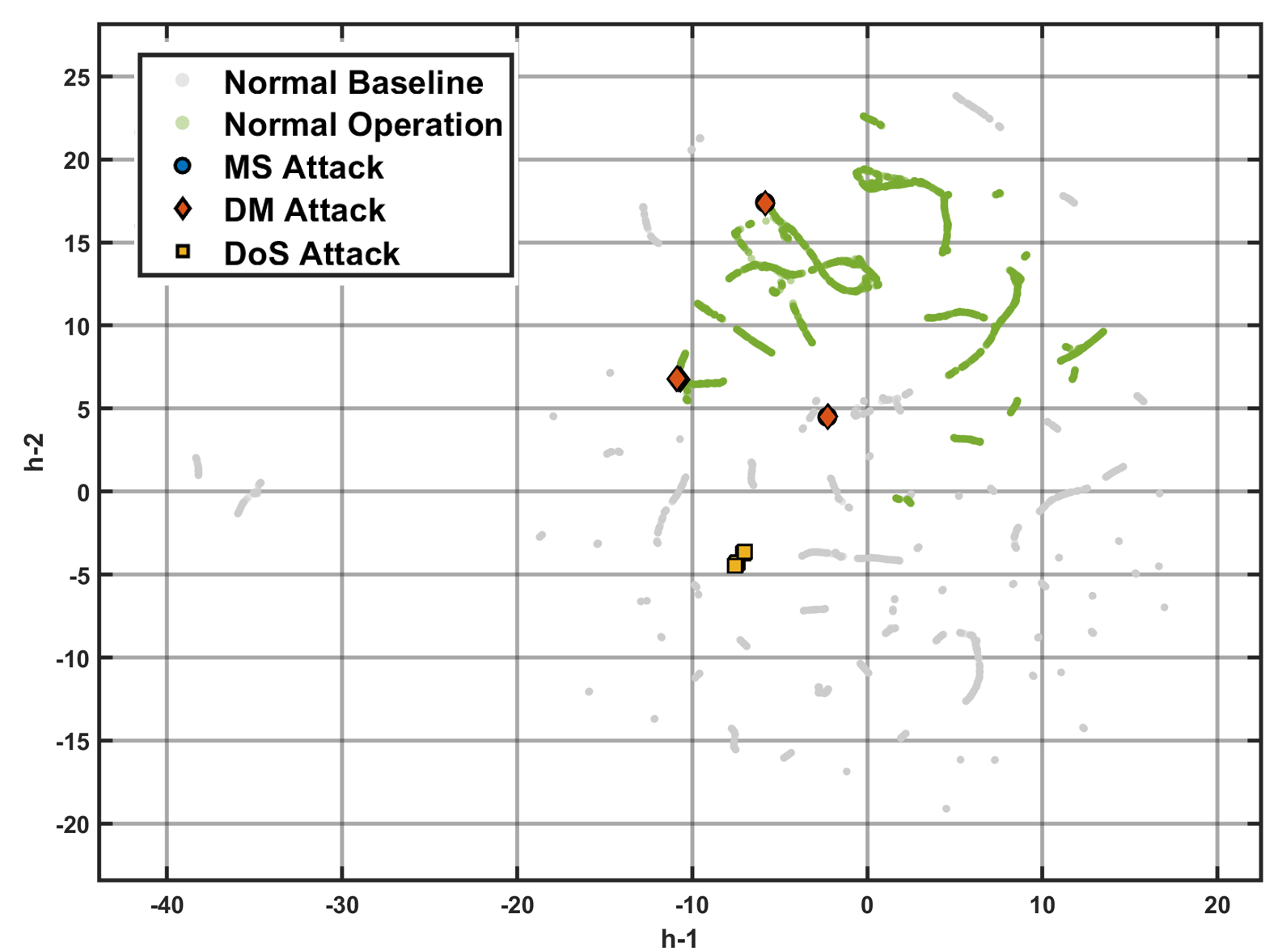} &
        \includegraphics[width=0.23\textwidth]{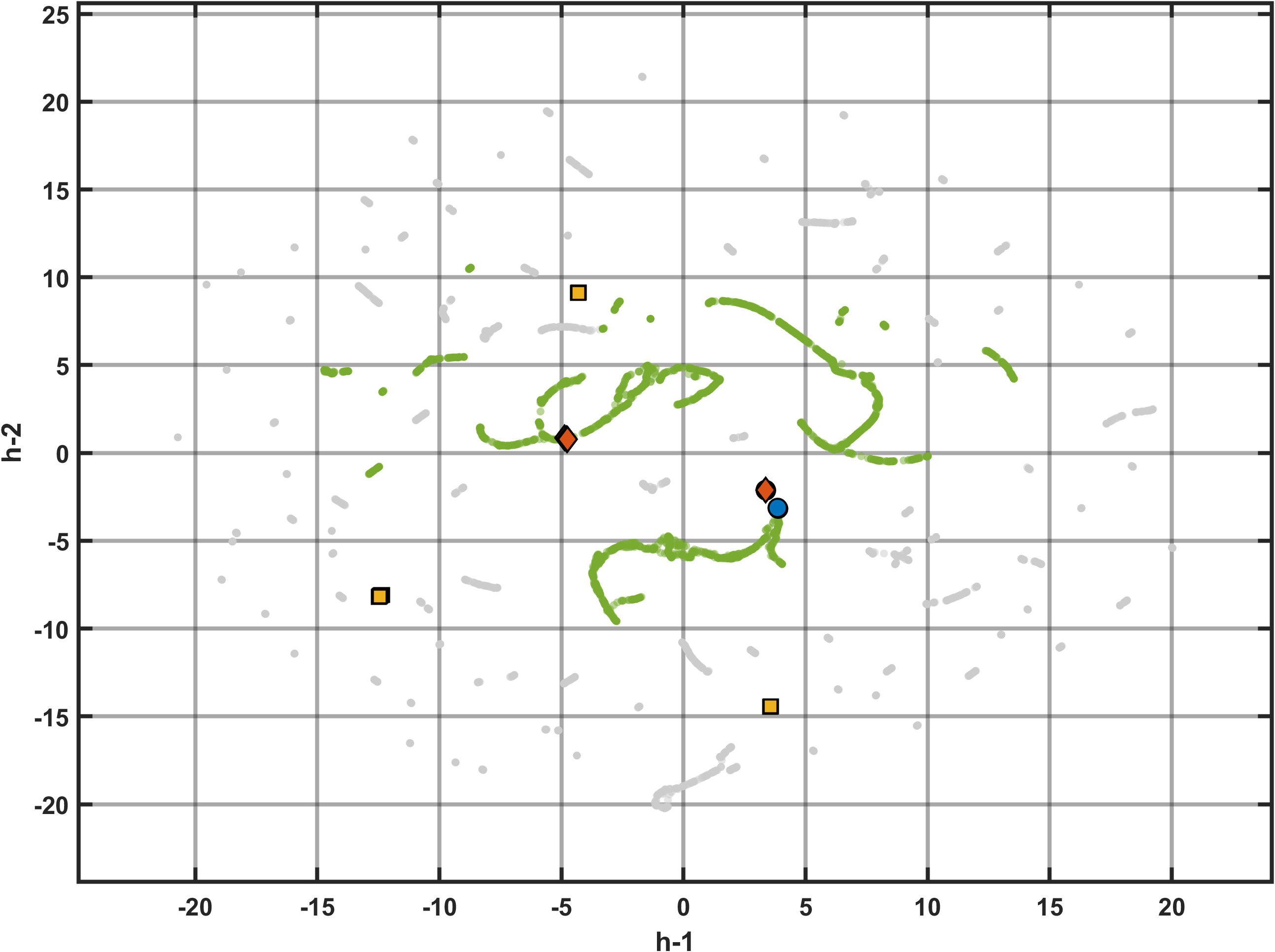} &
        \includegraphics[width=0.23\textwidth]{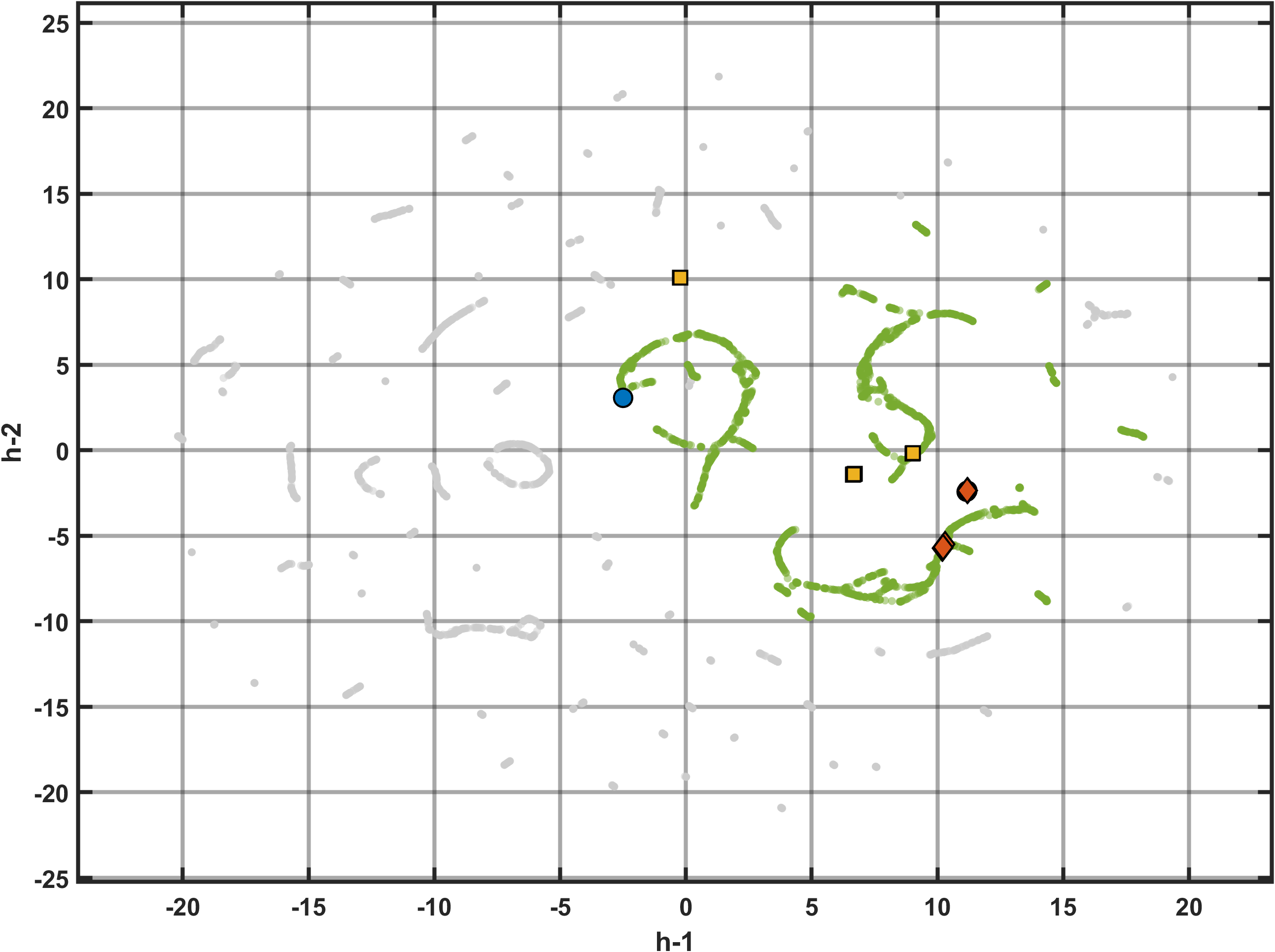} &
        \includegraphics[width=0.23\textwidth]{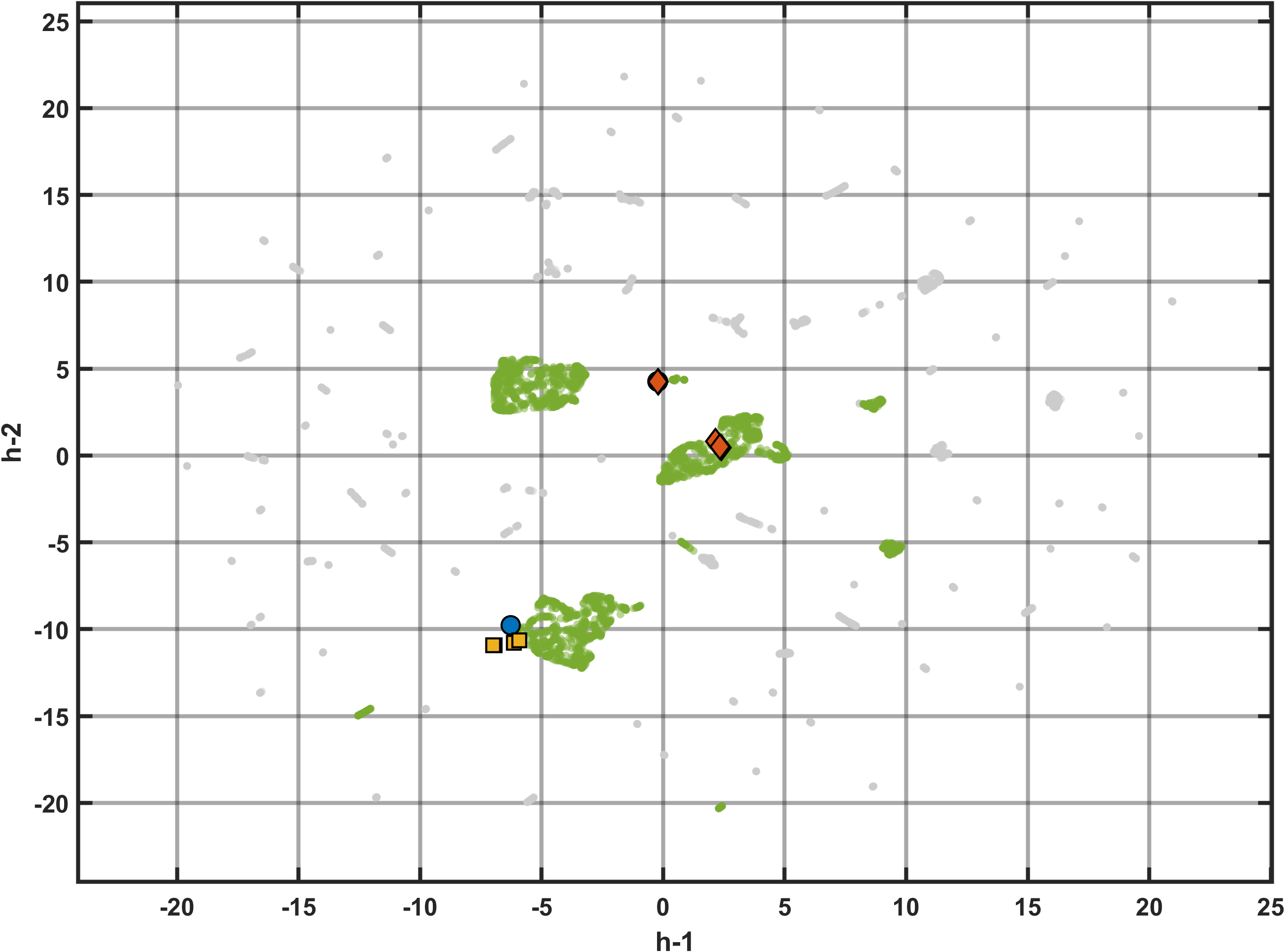} \\
        (e) $T_w = 0.1$ s & (f) $T_w = 0.5$ s & (g) $T_w = 1.0$ s & (h) $T_w = 3.0$ s \\
    \end{tabular}
    \caption{Visualization of the \textbf{Raw Temporal Feature Space}.}
    \label{fig:raw_space}
\end{figure*}
We use UMAP to examine the topology of the raw feature space in order to obtain information about the intrinsic structure of the features before model training. In this context, topological characterization refers to the neighborhood relationships and manifold structure induced by the distribution of the data in the high-dimensional feature space.

Fig. \ref{fig:raw_space} shows the two-dimensional UMAP embeddings obtained from the sequence (SEQ) and temporal (TEMP) feature sets for different time window lengths $T_w$. Gray points correspond to the real operational baseline used later for training, while colored points represent samples from the laboratory dataset, including both normal operation and attack scenarios.

The two feature domains show a distinct structural difference. The SEQ feature space forms compact and well-defined manifolds across all window sizes. Deterministic IEC 61850 semantics governs sequence features, therefore this behavior is expected. Normally, counters like \texttt{stNum} and \texttt{sqNum} evolve in accordance with strict constraints. As a result, normal traffic collapses into a dense region of the manifold, while most deviations appears in a different region. In contrast, the TEMP feature space has a more dispersed structure. Temporal features are inherently affected by network jitter, background traffic, and timing fluctuations, which introduce higher variability even during normal operation. Consequently, the corresponding manifold is less compact, with a weaker separation between normal and attack traffic.

The influence of the time window length is also evident. Very short windows ($T_w = 0.1$ s) result in fragmented embeddings that lack global coherence but capture fine-grained dynamics, especially in the TEMP feature space. Intermediate windows ($T_w =0.5$ s and $T_w = 1.0$ s) produce more structured manifolds, enhancing separability while preserving temporal detail. For longer windows ($T_w = 3$ s), aggregation effects dominate, leading to a partial loss of discriminative structure.

\subsection{Quantitative Performance Analysis}

\begin{table*}[h!]
\centering
\caption{Detection Performance by Attack Type and Time Window ($T_w$).}
\label{tab:window_sensitivity}
\setlength{\tabcolsep}{5pt}
\renewcommand{\arraystretch}{1.3}
\begin{tabular}{c|c|l||cccc|ccc}
\hline
\textbf{Window} & \textbf{Attack} & \textbf{Model} &
\textbf{TP} & \textbf{FP} & \textbf{TN} & \textbf{FN} &
\textbf{Recall} & \textbf{Spec.} & \textbf{F1} \\
\hline\hline

\multirow{6}{*}{\textbf{0.1 s}}
& \multirow{2}{*}{\textbf{DM}}
    & Semantic (SEQ)   & 11 &   6 & 95389 & 0 & 1.00    & 0.99994 & 0.78571 \\
&   & Temporal (TEMP) &  9 & 279 & 95116 & 2 & 0.81818 & 0.99708 & 0.06020 \\
\cline{2-10}

& \multirow{2}{*}{\textbf{MS}}
    & Semantic (SEQ)   &  7 &   7 & 127193 & 0 & 1.00    & 0.99994 & 0.66667 \\
&   & Temporal (TEMP) &  4 & 380 & 126820 & 3 & 0.57143 & 0.99701 & 0.02046 \\
\cline{2-10}

& \multirow{2}{*}{\textbf{DoS}}
    & Semantic (SEQ)   & 41 & 195 & 31789 & 0 & 1.00    & 0.99390 & 0.29603 \\
&   & Temporal (TEMP) & 41 & 287 & 31697 & 0 & 1.00    & 0.99103 & 0.22222 \\
\hline\hline

\multirow{6}{*}{\textbf{0.5 s}}
& \multirow{2}{*}{\textbf{DM}}
    & Semantic (SEQ)   &  7 &   6 & 95388 & 4 & 0.63636 & 0.99994 & 0.58333 \\
&   & Temporal (TEMP) &  6 & 718 & 94676 & 5 & 0.54545 & 0.99247 & 0.01633 \\
\cline{2-10}

& \multirow{2}{*}{\textbf{MS}}
    & Semantic (SEQ)   &  7 &   4 & 127196 & 0 & 1.00    & 0.99997 & 0.77778 \\
&   & Temporal (TEMP) &  5 & 1074 & 126120 & 2 & 0.71429 & 0.99156 & 0.00921 \\
\cline{2-10}

& \multirow{2}{*}{\textbf{DoS}}
    & Semantic (SEQ)   & 48 &   1 & 31789 & 0 & 1.00    & 0.99997 & 0.98969 \\
&   & Temporal (TEMP) & 48 & 239 & 31551 & 0 & 1.00    & 0.99248 & 0.28657 \\
\hline\hline

\multirow{6}{*}{\textbf{1.0 s}}
& \multirow{2}{*}{\textbf{DM}}
    & Semantic (SEQ)   &  7 &   0 & 95381 & 4 & 0.63636 & 1.00    & 0.77778 \\
&   & Temporal (TEMP) &  7 & 618 & 94763 & 4 & 0.63636 & 0.99352 & 0.02201 \\
\cline{2-10}

& \multirow{2}{*}{\textbf{MS}}
    & Semantic (SEQ)   &  4 &   3 & 127180 & 3 & 0.57143 & 0.99998 & 0.57143 \\
&   & Temporal (TEMP) &  5 & 880 & 126300 & 2 & 0.71429 & 0.99308 & 0.01121 \\
\cline{2-10}

& \multirow{2}{*}{\textbf{DoS}}
    & Semantic (SEQ)   & 26 &   0 & 31790 & 0 & 1.00    & 1.00    & 1.00 \\
&   & Temporal (TEMP) & 26 & 206 & 31584 & 0 & 1.00    & 0.99352 & 0.20155 \\
\hline\hline

\multirow{6}{*}{\textbf{3.0 s}}
& \multirow{2}{*}{\textbf{DM}}
    & Semantic (SEQ)   &  5 &   0 & 33774 & 4 & 0.55556 & 1.00    & 0.71429 \\
&   & Temporal (TEMP) &  6 & 231 & 33543 & 3 & 0.66667 & 0.99316 & 0.04878 \\
\cline{2-10}

& \multirow{2}{*}{\textbf{MS}}
    & Semantic (SEQ)   &  7 &   2 & 44984 & 0 & 1.00    & 0.99996 & 0.87500 \\
&   & Temporal (TEMP) &  6 & 372 & 44614 & 1 & 0.85714 & 0.99173 & 0.03117 \\
\cline{2-10}

& \multirow{2}{*}{\textbf{DoS}}
    & Semantic (SEQ)   & 10 &   1 & 11238 & 0 & 1.00    & 0.99991 & 0.95238 \\
&   & Temporal (TEMP) & 10 &  78 & 11161 & 0 & 1.00    & 0.99306 & 0.20408 \\
\hline

\end{tabular}
\end{table*}

\begin{figure*}[h]
    \centering
    \begin{tabular}{cccc}
        \multicolumn{4}{c}{\textbf{Semantic latent space (SEQ)}} \\
        \includegraphics[width=0.23\textwidth]{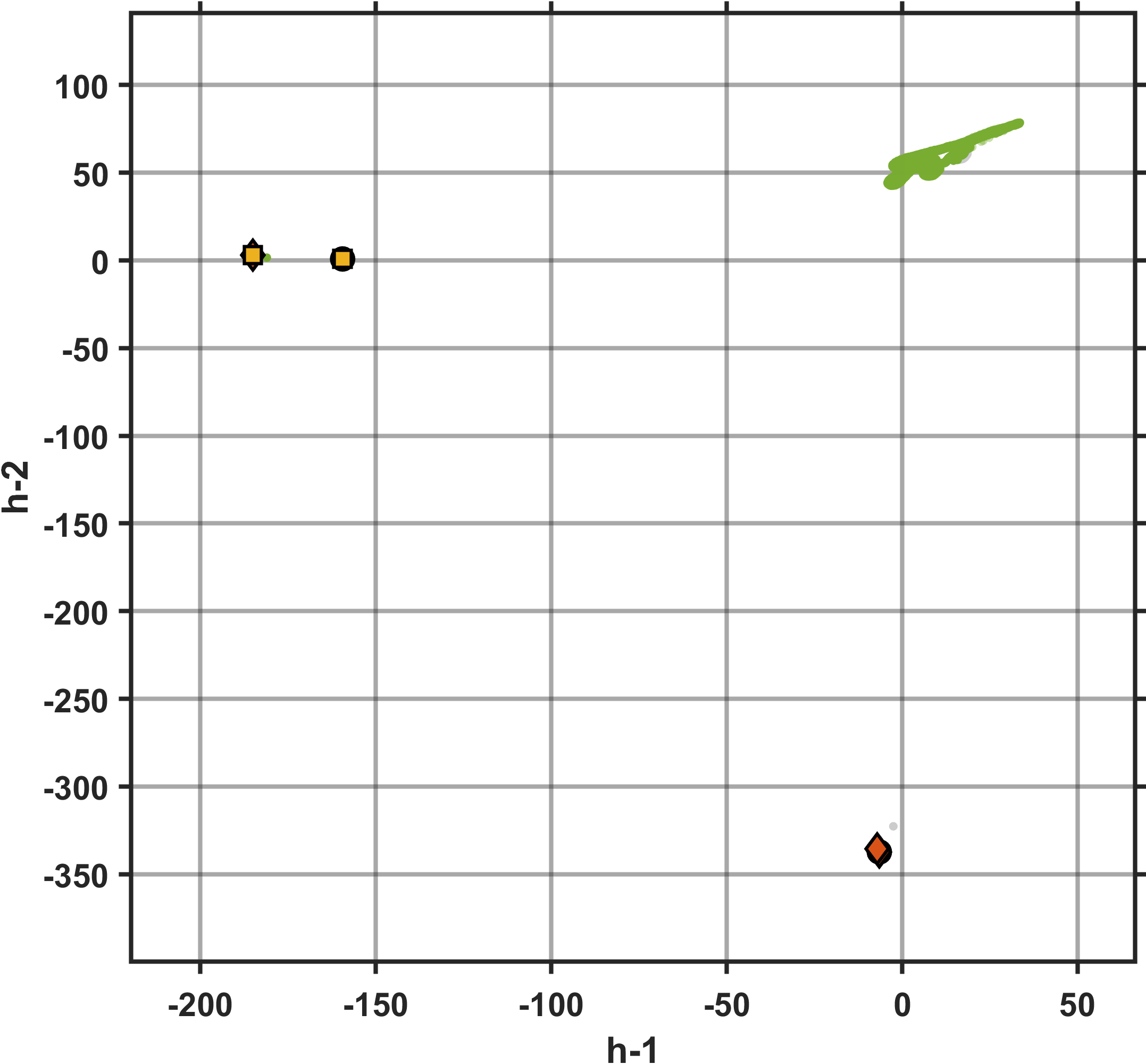} &
        \includegraphics[width=0.23\textwidth]{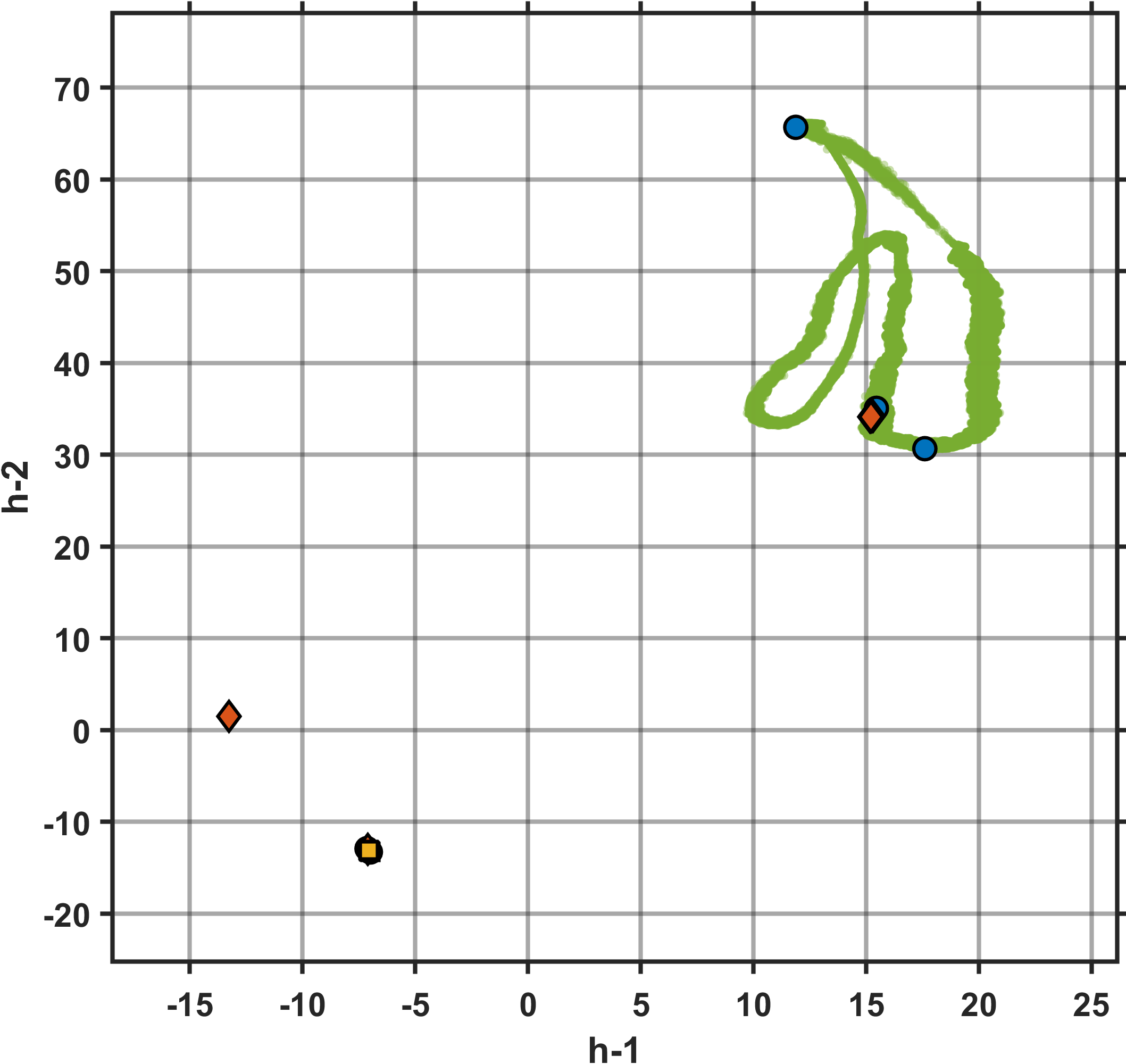} &
        \includegraphics[width=0.23\textwidth]{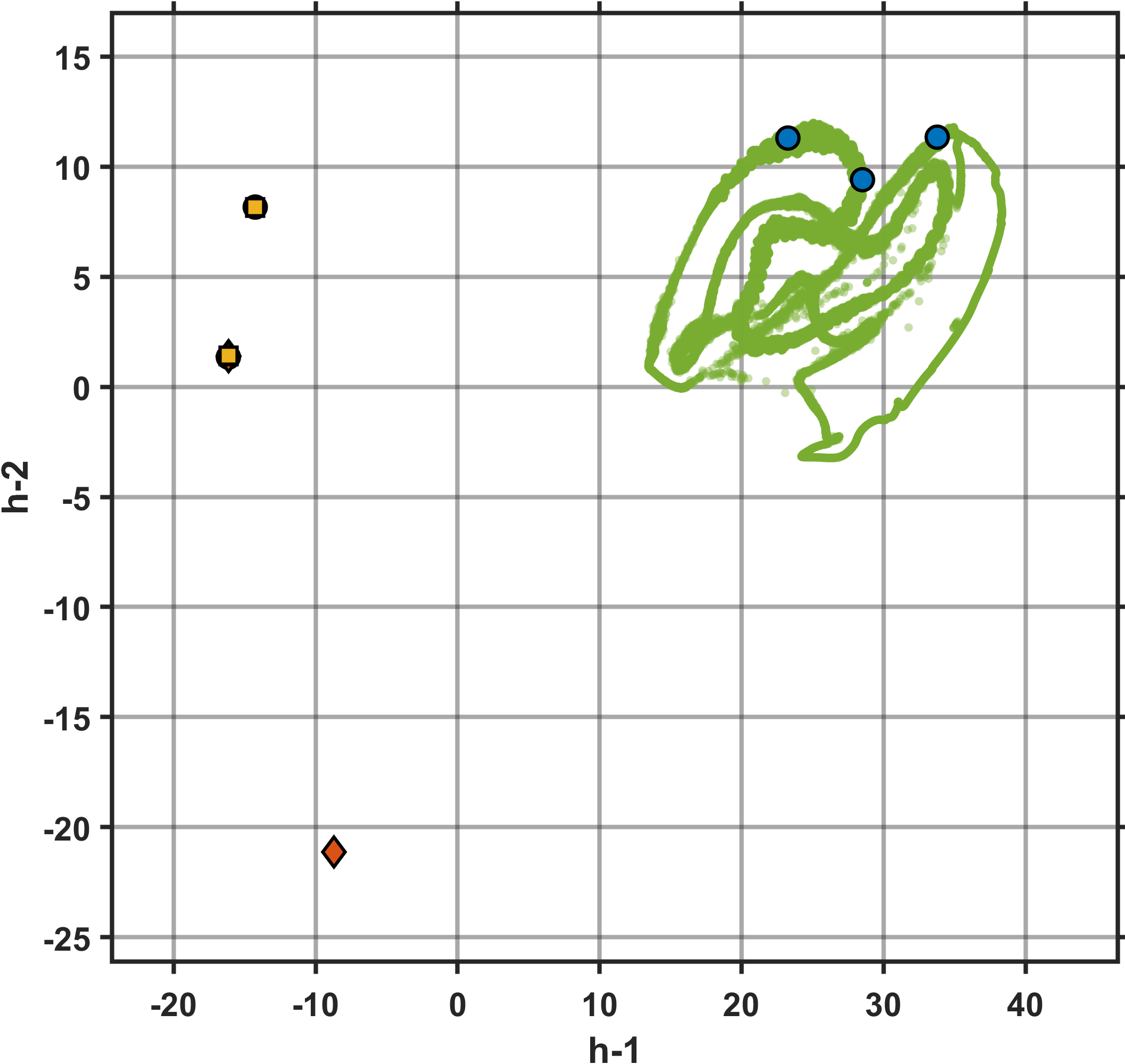} &
        \includegraphics[width=0.23\textwidth]{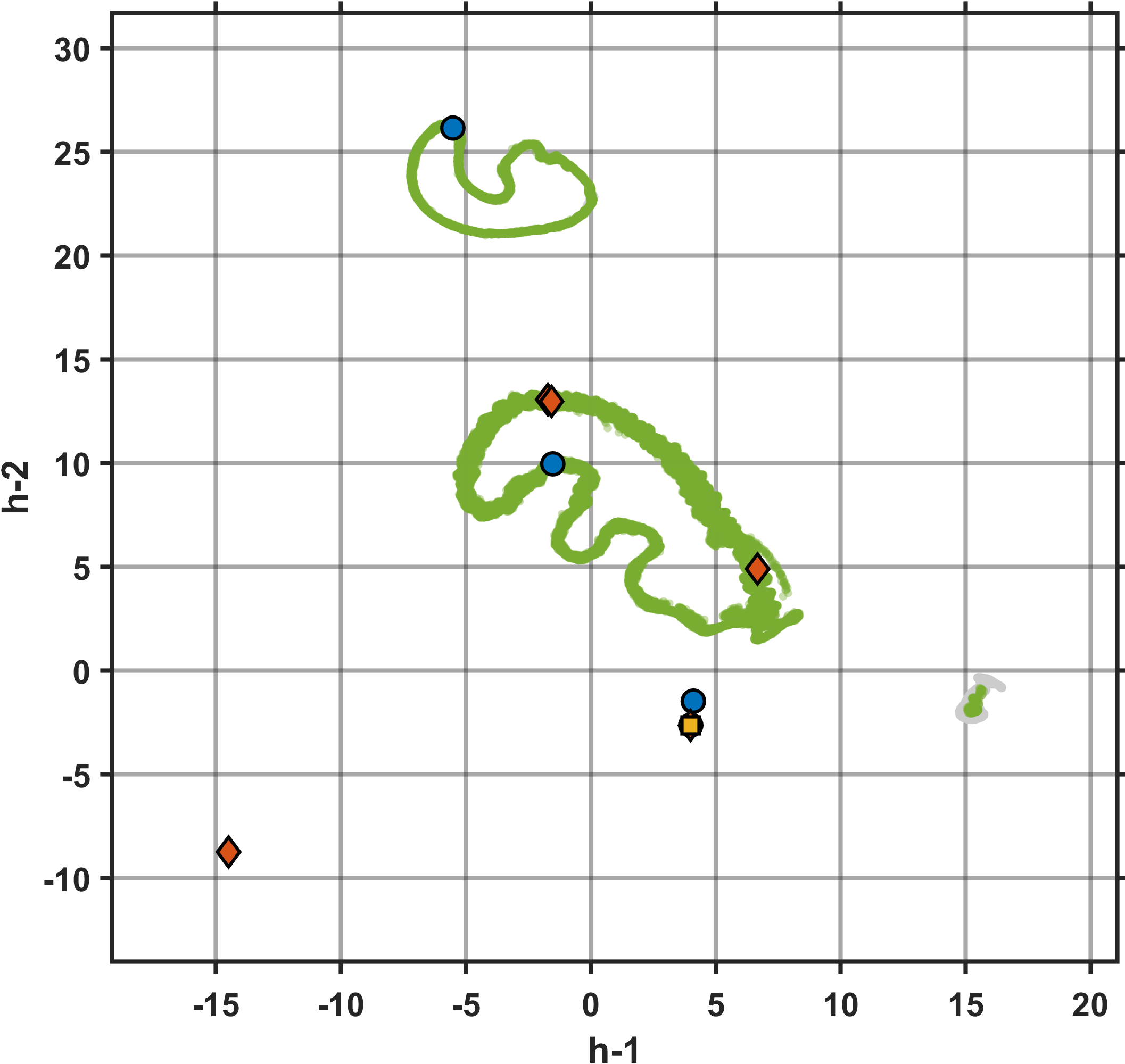} \\
        (a) $T_w = 0.1$ s & (b) $T_w = 0.5$ s & (c) $T_w = 1.0$ s & (d) $T_w = 3.0$ s \\
        [1.5ex]
        \multicolumn{4}{c}{\textbf{Temporal latent space (TEMP)}} \\
        \includegraphics[width=0.23\textwidth]{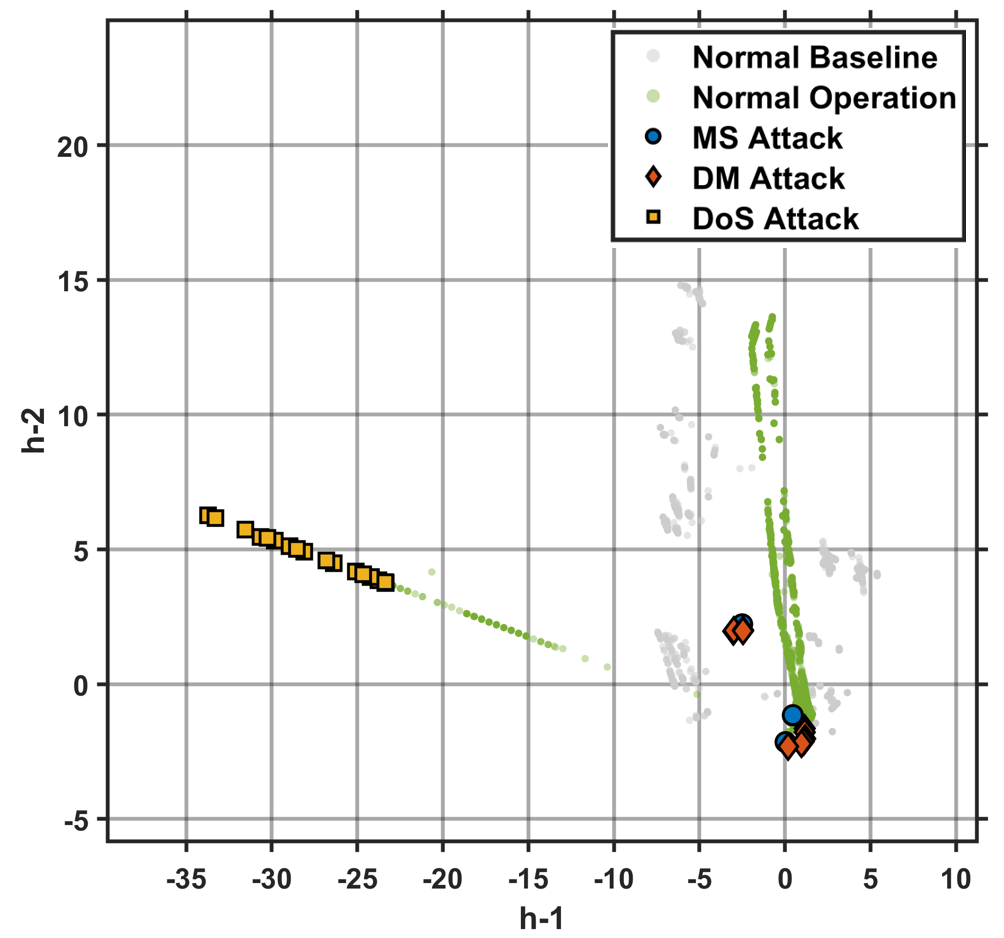} &
        \includegraphics[width=0.23\textwidth]{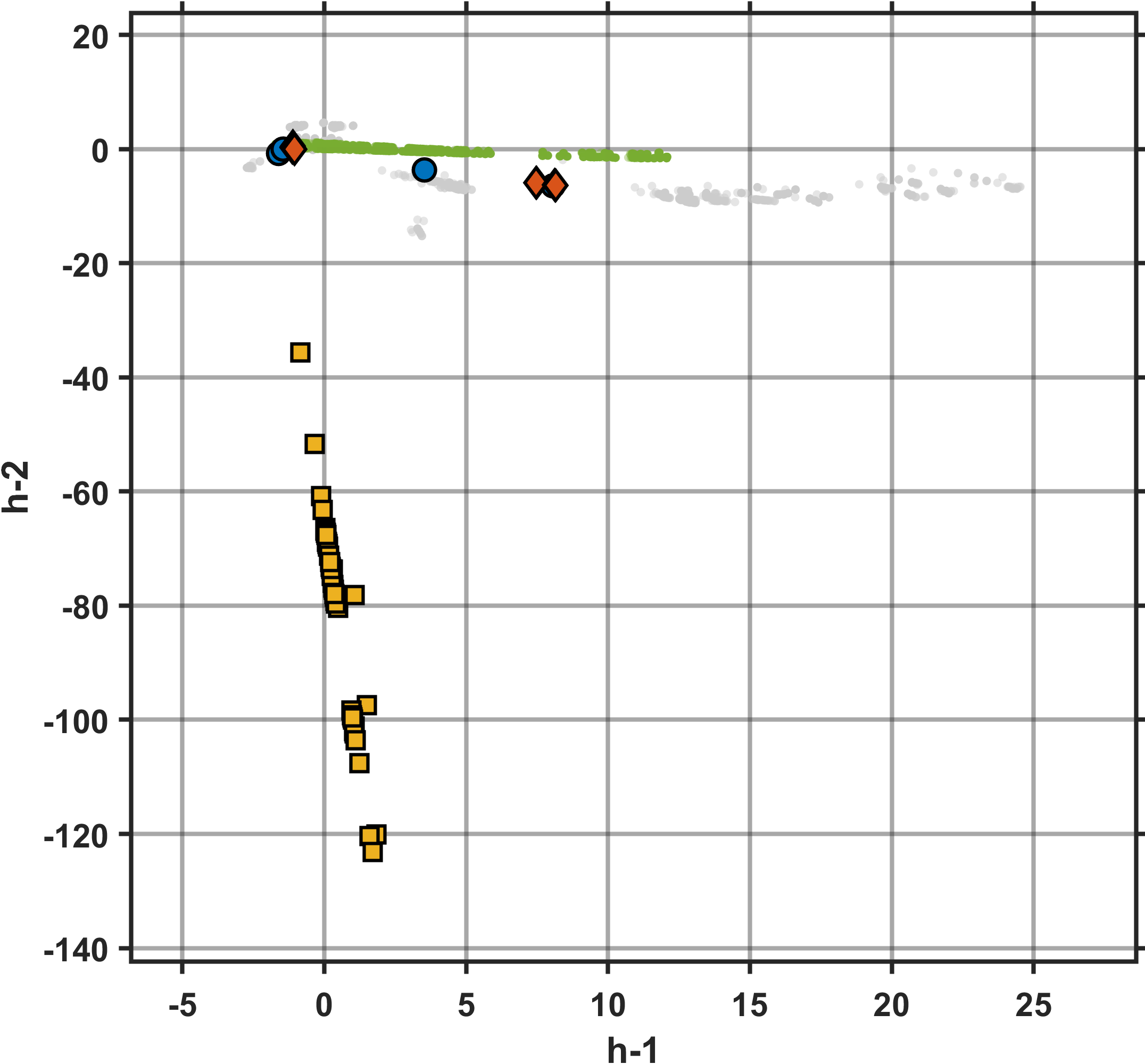} &
        \includegraphics[width=0.23\textwidth]{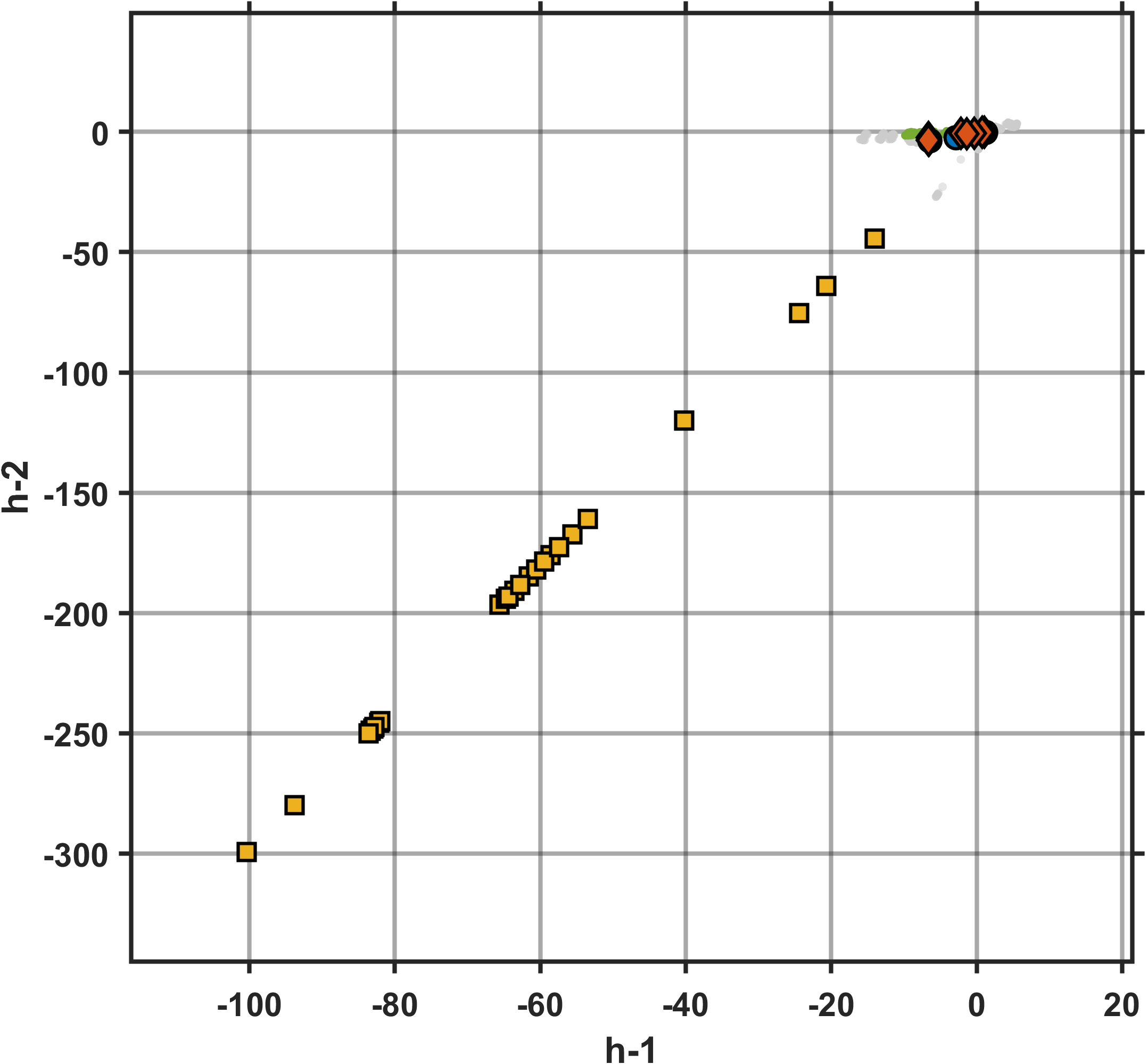} &
        \includegraphics[width=0.23\textwidth]{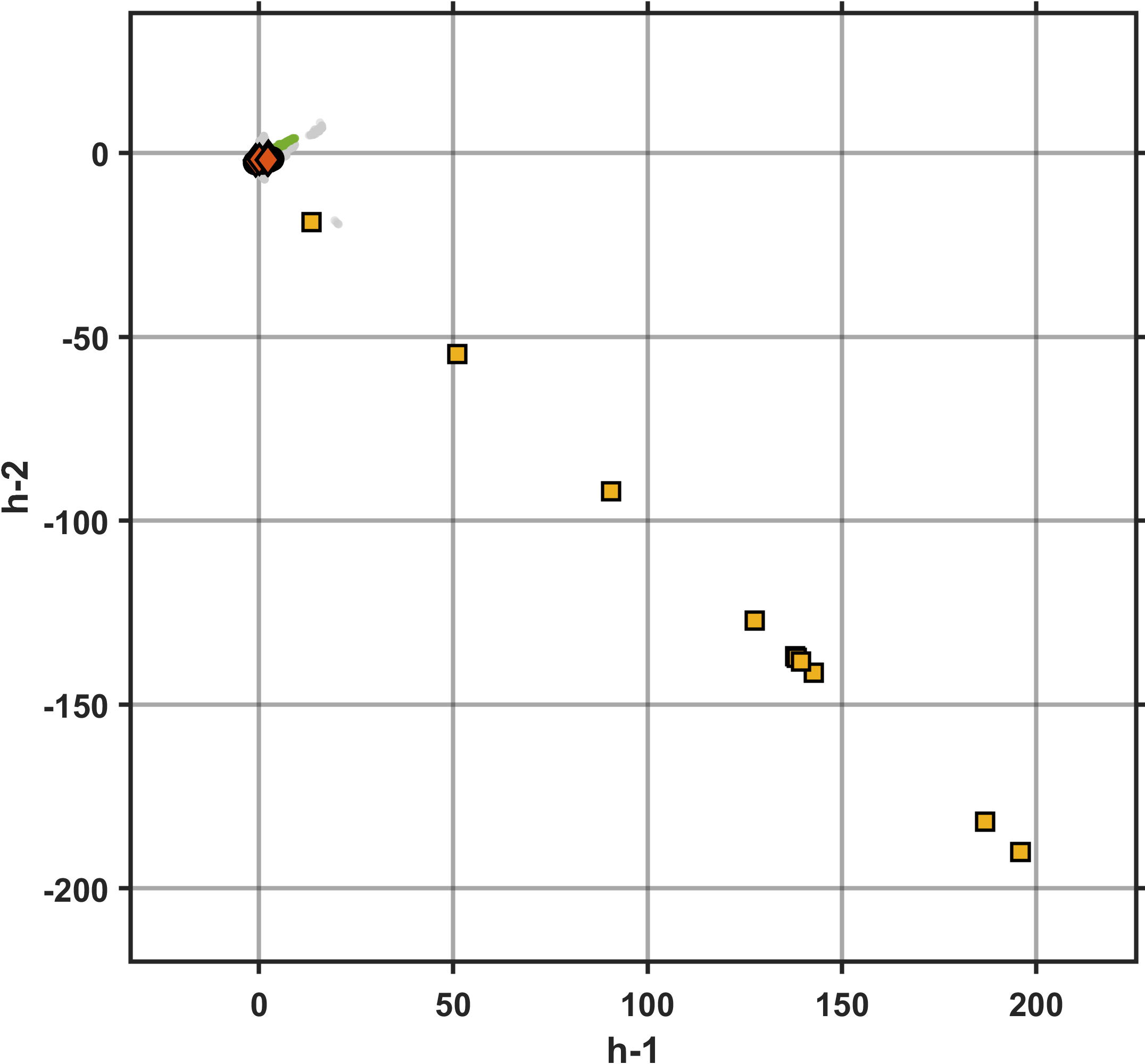} \\
        (e) $T_w = 0.1$ s & (f) $T_w = 0.5$ s & (g) $T_w = 1.0$ s & (h) $T_w = 3.0$ s \\
    \end{tabular}
    \caption{Visualization of the AE latent spaces. Direct projection of the bottleneck representations learned by the AEs for different time window lengths.}
    \label{fig:latent_spaces}
\end{figure*}

The previous section showed that the raw feature space exhibits different levels of separability depending on the feature domain and the selected time window. In several scenarios, anomalous behavior partially overlaps with normal traffic. This observation motivates the use of reconstruction-based anomaly detection approaches, such as AEs.

Table \ref{tab:window_sensitivity} reports the detection performance of the proposed AE for each attack type and time window $T_w$, using the dynamic thresholds derived via EVT. These thresholds allow the model to operate conservatively, ensuring high sensitivity to anomalous behavior while maintaining strong control over false positives.

Across all window sizes, the semantic (SEQ) AE consistently achieves high recall, often reaching 1.00 for MS, DM and DoS attacks. This confirms that interruptions of sequencing rules produce reconstruction errors that are reliably separable from normal behavior. In contrast, the temporal (TEMP) AE shows lower precision and F1-scores, particularly for short and intermediate windows. Temporal anomalies typically appear as gradual deviations rather than abrupt changes, which leads to a higher number of false positives despite good recall. Nevertheless, the proportion of false positives remains below 5\% relative to the total traffic volume.

The influence of the time window length is also evident. Intermediate windows ($T_w =0.5$ s and $T_w = 1.0$ s) provide the best trade-off between temporal resolution and stability. For these window sizes, the SEQ AE achieves near-perfect detection of DoS attacks and robust performance for MS attacks, while maintaining high specificity. Very short windows ($T_w = 0.1$ s) increase sensitivity to transient effects but also amplify noise, particularly in the temporal domain. On the contrary, longer windows ($T_w = 3.0$ s) smooth short-lived anomalies, reducing sensitivity for certain attack types.

\subsection{Manifold Structure and Topological Characterization of Latent Space}

Fig. \ref{fig:latent_spaces} analyzes the topology of the learned latent representations produced by the AE bottlenecks. By imposing an extreme compression of the input space, the AE are forced to retain only the most informative factors of variation associated with legitimate GOOSE behavior. 

The SEQ latent space shows a clear and stable structure, where normal operation, including both real-world baseline traffic and synthetic normal samples, concentrates into a single dense region. This characteristic provides evidence of strong semantic invariance in GOOSE protocol for normal traffic. Most attack samples appear separated from this normal manifold, occupying distinct regions of the latent space. Differences in the dispersion and direction of attack manifolds can be attributed to the specific nature of each attack type and the time window length, with longer windows producing smoother and more structured trajectories. In contrast, the TEMP latent space remains more dispersed, mirroring the higher variability inherent to timing-related features even under normal operation. Despite this dispersion, DoS attacks consistently emerge as well-separated structures, particularly for intermediate and long time windows. This separation is driven by the sustained rate and volume changes introduced by flooding, which create persistent deviations from normal temporal behavior and can therefore be detected even within a noisy temporal space.
 
\begin{figure*}[h]
\begin{tabular}{cc}
\begin{minipage}[t]{0.5\linewidth}
    \includegraphics[width=\linewidth]{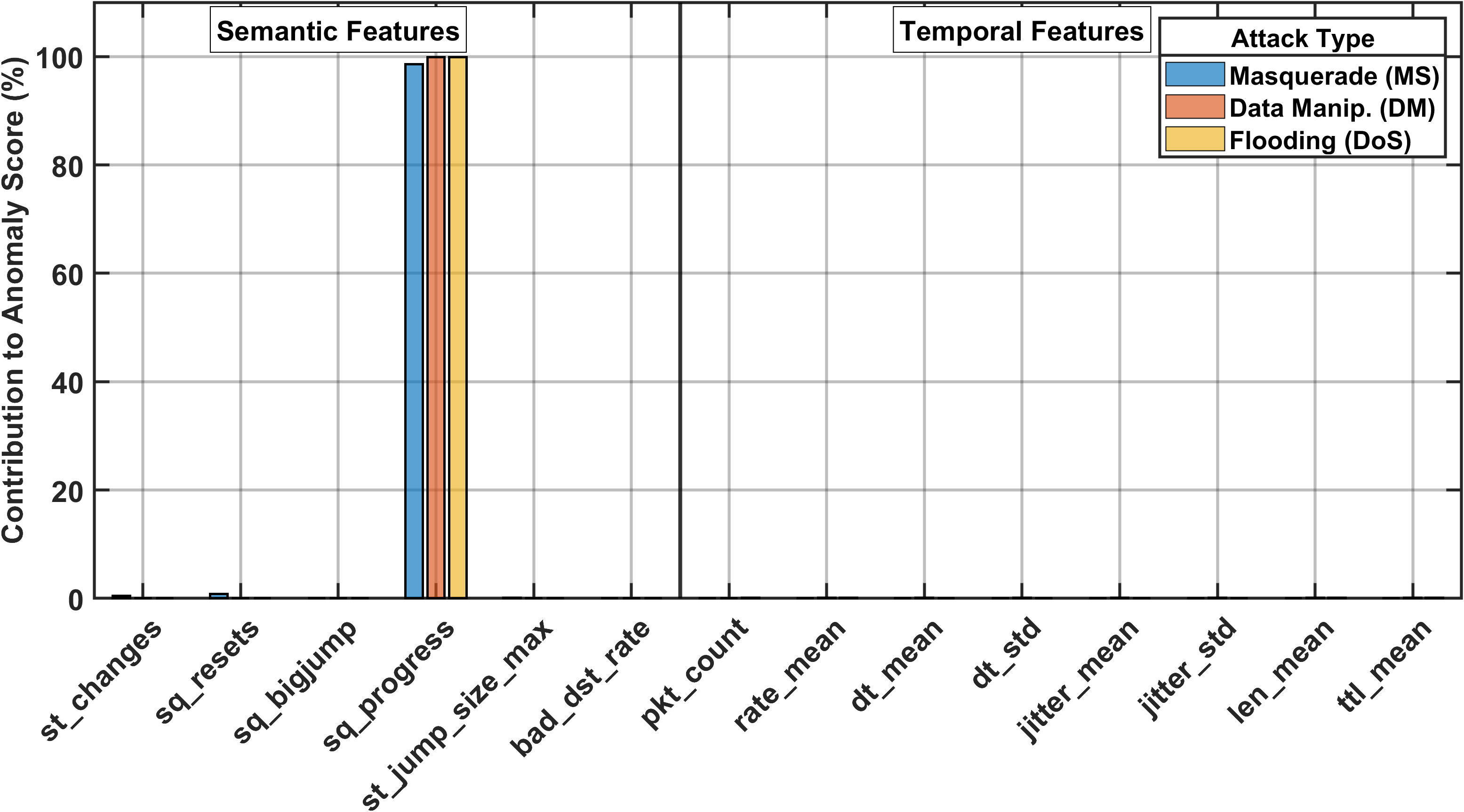}
\end{minipage}
&
\begin{minipage}[t]{0.5\linewidth}
    \includegraphics[width=\linewidth]{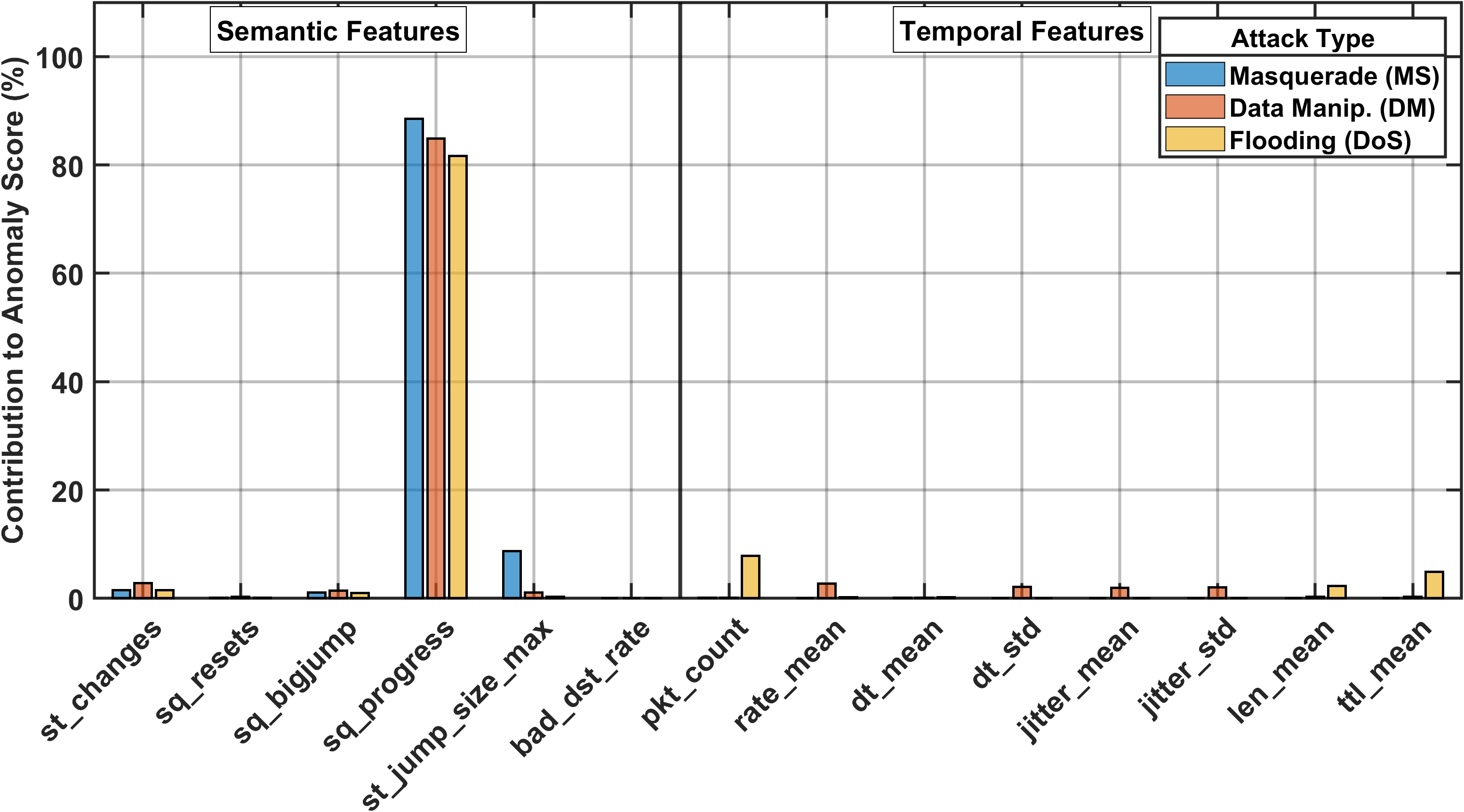}
\end{minipage}
\\

(a) $T_w = 0.1$ s & (b) $T_w = 0.5$ s 
\\
\\
\begin{minipage}[t]{0.5\linewidth}
    \includegraphics[width=\linewidth]{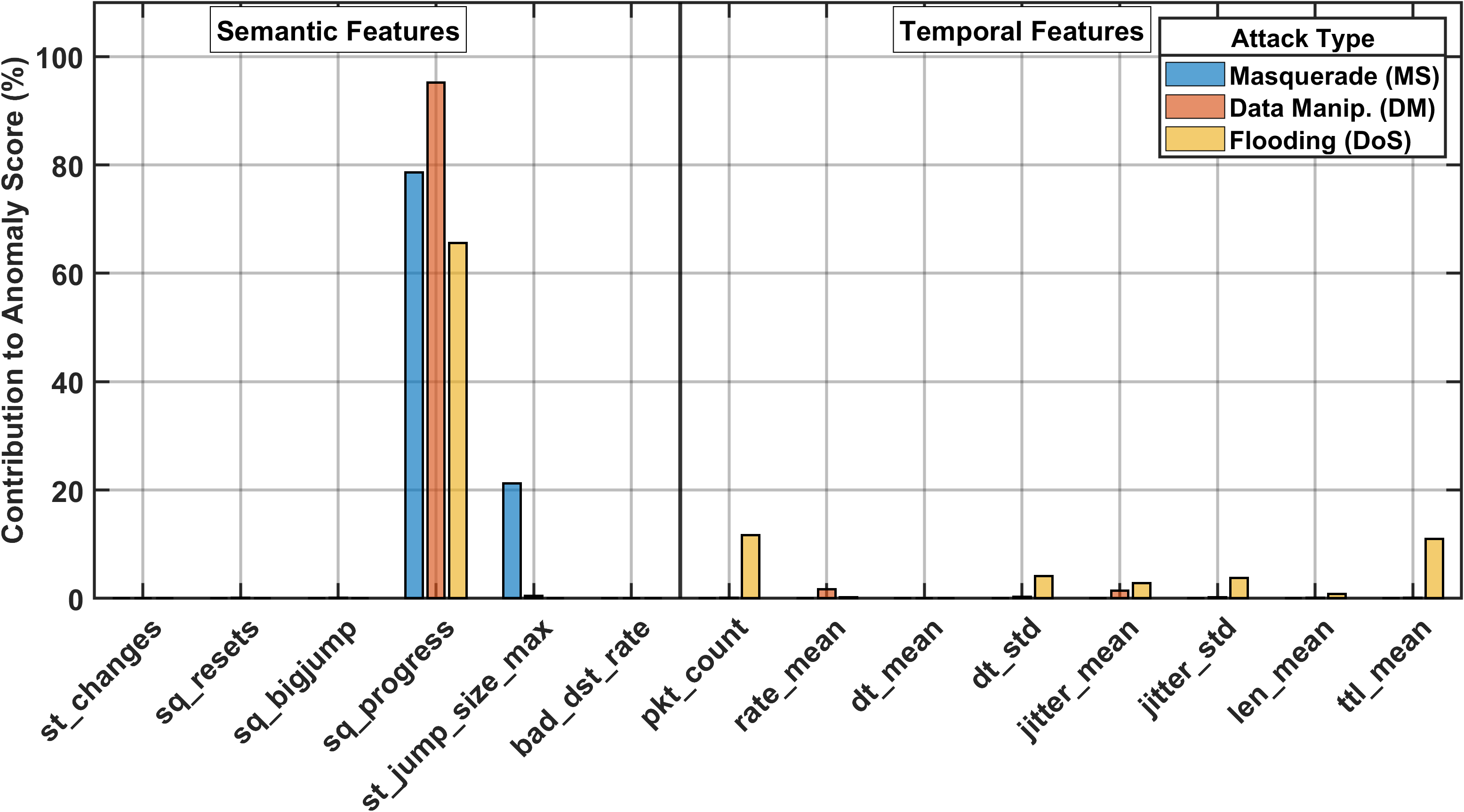}
\end{minipage}
&
\begin{minipage}[t]{0.5\linewidth}
    \includegraphics[width=\linewidth]{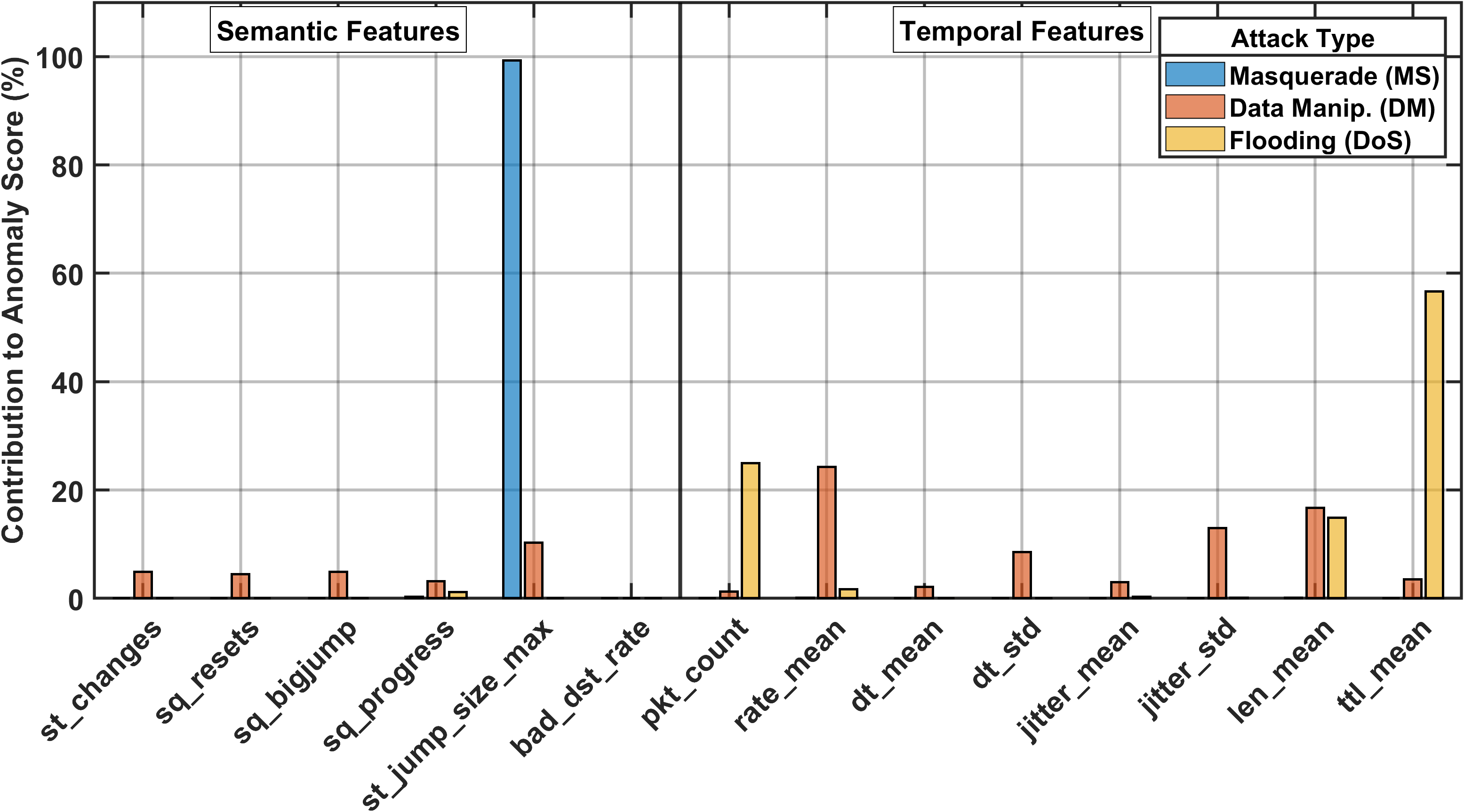}
\end{minipage}

\\
(c) $T_w = 1.0$ s & (d) $T_w = 3.0$ s \\
\end{tabular}
\caption{Feature contribution to the anomaly score for different attack types and time window lengths.}
\label{fig:XAIFeatures}
\end{figure*}

\subsection{Feature Importance Attribution}

To support interpretation of detected anomalies, Fig. \ref{fig:XAIFeatures} reports the normalized per-feature contribution to the anomaly score, computed from the squared reconstruction error of each input variable. Therefore, highlighting the features most responsible for the detection decision.

Across all time windows, the attacks MS and DM are dominated by sequence-based features, most notably \textit{sq\_progress} and \textit{st\_jump\_size\_max}. This behavior reflects deviations of the deterministic sequencing rules defined by the GOOSE protocol, where sequence numbers are expected to evolve continuously and state numbers change only in response to legitimate events. Suppression and manipulation attacks disrupt these rules, which directly affect the continuity of sequence progression and the magnitude of state changes. The stability of their contribution across different window lengths further indicates that semantic deviations are robustly preserved even under temporal aggregation. On the contrary, temporal features play a secondary role for MS and DM attacks but become increasingly relevant for DoS scenarios, particularly at larger window sizes ($T_w = 1.0$ s and $T_w = 3.0$ s). In these cases, volumetric and timing-related variables such as \textit{pkt\_count} and rate-based descriptors (\textit{ttl\_mean}) contribute substantially to the anomaly score, reflecting the sustained traffic patterns introduced by flooding attacks.

\section{Discussion}
\label{sec:discussion}

This work investigates anomaly detection in IEC~61850 GOOSE networks using a multi-view approach that separates semantic and temporal features. The results show that these two aspects of GOOSE behavior exhibit fundamentally different characteristics and therefore benefit from distinct modeling strategies.

The asymmetric AE design plays a central role in the results. By modeling sequence-based and temporal features independently, the model is able to detect most of the anomalies. The SEQ AE learns the normal behavior of GOOSE sequence counters and reconstructs legitimate traffic accurately, while deviations introduced by attacks lead to higher reconstruction errors. As a result, the SEQ AE consistently achieves high recall with very few, if any, false positives. In contrast, the TEMP AE operates on features that are inherently more variable due to network jitter, background traffic, and timing fluctuations. This variability explains the higher number of false positives observed in the temporal view, particularly for short and intermediate windows. Nevertheless, all evaluated attacks are correctly detected, and the proportion of false positives remains below 5\% of the total traffic volume, which is acceptable in realistic substation environments where attacks are rare. Dynamic thresholding based on EVT also helps stabilize the detection process. Instead of using fixed thresholds, EVT focuses on the largest reconstruction errors observed during normal operation and uses them to define adaptive thresholds. This makes it possible to detect attacks reliably while keeping the number of false alarms at a manageable level, even under severe class imbalance and heterogeneous traffic conditions.

The learned latent representations provide additional insight into these results. For the SEQ AE, both real operational traffic and synthetic normal traffic are mapped to the same compact region of the latent space, while attack samples are clearly separated. A similar alignment between real and synthetic normal traffic is also observed in the TEMP AE. This consistent behavior across datasets indicates that the model learns protocol-level characteristics rather than dataset-specific patterns. As a result, the proposed approach generalizes well to different environments without requiring retraining or manual tuning. However, the TEMP AE shows a more dispersed latent structure overall, reflecting the higher variability of timing-related features. Despite this dispersion, DoS attacks remain distinguishable due to their sustained impact on traffic volume and transmission rate.

Finally, the results also highlight the importance of time window selection. Intermediate window lengths between 0.5 and 1.0~s provide the most favorable balance between sensitivity to short-lived anomalies and robustness to noise, whereas very short and very long windows tend to emphasize transient fluctuations or temporal averaging effects, respectively.

\section{Conclusion}
\label{sec:conclusion}

This paper presented a robust and explainable multi-view anomaly detection framework for IEC~61850 GOOSE networks. By explicitly separating semantic integrity from temporal availability, the proposed approach captures complementary aspects of GOOSE behavior that are difficult to model using single-model detectors. The combination of asymmetric AEs, dynamic thresholding, and feature-level attribution enables accurate detection of MS, DM, and DoS using only normal operational traffic for training. Experimental results on real substation data and a public attack dataset demonstrate strong generalization across environments, high attack detection rates under extreme class imbalance, and interpretable detection outcomes. These characteristics make the proposed framework suitable for anomaly detection in operational settings, including scenarios involving previously unseen attacks, where labeled attack data are scarce and explainability is a key requirement.


\section*{Acknowledgments}
This research received funding from the CyberFold project within the European Union’s NextGenerationEU program (Recovery, Transformation, and Resilience Plan), with project management provided by the Instituto Nacional de Ciberseguridad de España (INCIBE) under reference ETD202300129. Additional funding was provided by the Autonomous Community of Madrid through the ELLIS Madrid Node. Also, partial funding was provided by the project PID2022-140786NB-C32 (LATENTIA), funded by the Spanish Ministry of Science and Innovation (AEI/10.13039/501100011033).

\bibliographystyle{unsrtnat}
\bibliography{mybibfile}

\end{document}